# Semiconductor Room-Temperature Maser

Andreas Gottscholl[1,2,a)], Maximilian Wagenhöfer[1], Valentin Baianov[1], Emilian Eisermann[1], Vladimir Dyakonov[1], Andreas Sperlich[1,b)]

[1]Experimental Physics 6 and Würzburg-Dresden Cluster of Excellence ctd.qmat, Julius-Maximilians-Universität Würzburg, 97074 Würzburg, Germany

[2]NASA Jet Propulsion Laboratory, California Institute of Technology, Pasadena, California, USA

[a)] Electronic mail: gottscholl.andreas@gmail.com

[b)] Electronic mail: andreas.sperlich@uni-wuerzburg.de

## ABSTRACT

We report the first demonstration of a semiconductor maser based on silicon vacancies ($V_{Si}$) in 4H-silicon carbide (SiC). Using an active feedback loop, we enhance the resonator's quality factor, enabling continuous-wave maser operation even above room temperature. We analyzed the SiC maser as a high-performance preamplifier, with measured gain exceeding 10 dB at 110 K and simulations suggesting potential amplification beyond 30 dB. Leveraging the small zero-field splitting of $V_{Si}$, the device can also function as an optically pumped microwave photon absorber, reducing the resonator's mode temperature by 40 K relative to the environment. Additionally, the maser's ultranarrow linewidth supports highly sensitive magnetometry, achieving a nine-order-of-magnitude improvement in contrast-to-linewidth ratio over electrical and optical detection of magnetic resonance. This results in an estimated magnetic field sensitivity of 20 pT/√Hz at room-temperature based on the relative intensity noise of the excitation laser. These results underscore the potential of SiC to reshape room-temperature maser technologies, and lay the groundwork for future development of compact, electrically driven maser diodes.

## INTRODUCTION

Lasers have become an indispensable part of our everyday lives. From high-power lasers in manufacturing technology to high-precision lasers in metrology or monochromatic light sources for research, lasers and their enormous application potential are ubiquitous. [1,2] The fundamental process of stimulated emission based on a population inversion is however older than the laser and was previously investigated for microwave photons instead of optical ones [3,4] which is also known as a maser (microwave amplification of stimulated emission).

A broad application of a maser, however, is currently not predictable. So far masers were only used as low-noise amplifier and clock-standards depending on their linewidth and stability. This is owed to the very unfavorable operating conditions such as vacuum techniques for free-electron and atomic masers and cryogenic temperatures for solid state masers. [5-7] Therefore, the enormous (and so far largely unexplored) potential of masers has not yet been exploited.

Despite these challenges, recent advancements in spin-defect masers and other approaches, such as the exploitation of Floquet systems, [8] have opened new possibilities for overcoming these limitations. Among spin-defect masers, which promise fast and versatile application due to their comparatively simple design, there are two known systems so far: On the one hand Oxborrow et al presented the first room temperature maser using organic materials (e.g. pentacene-based) with a pulsed maser output at 1.45 GHz, which is currently developed towards a continuous wave (CW) output. [9-13] On the other hand, room temperature masers were demonstrated with $NV^-$ centers in diamond providing enhanced thermal and mechanical properties. [14-16] Due to its EPR-based (electron paramagnetic resonance) principle it operates at a X-band frequency of 9.2 GHz in a CW mode. In general, its frequency can be tuned to an arbitrary frequency by the external magnetic field and is only limited to the resonance frequency of the used resonator. Besides the existing maser systems, another promising candidate are silicon vacancies in 4H silicon carbide (SiC). [17-19] Their very small ZFS of only $2D/h = 70$ MHz makes the SiC maser attractive for applications within the entire communication band, since a small ZFS corresponds to a very broad frequency tunability, since population inversion of the maser transition is reached already for small bias magnetic fields. However, a fully operating maser has been illusive so far. SiC is known as a well-established semi-conductor material in industry and can be fabricated on wafer-scale. Thus, a maser based on this standard material could enable a new generation of spin-defect masers, serving not only as an exciting platform for cavity quantum dynamics but also as a gamechanger in everyday microwave applications, with the potential to revolutionize state-of-the-art preamplifiers, coherent microwave sources, and even enable unpredictable new applications such as highly sensitive quantum magnetometers.

## RESULTS

### Silicon Carbide as gain medium

The basic principle of a maser is identical to a conventional laser and is illustrated in Fig. 1a. [9,14] The maser consists of three major parts: 1) a gain material, 2) optical pumping in order to achieve a non-equilibrium condition (population inversion) and 3) a resonator in which the gain material is placed which enhances the interaction of microwaves and gain material. In the following we want to discuss the different components.

1) Gain material based on SiC: For the gain material we choose a wide-spread material which can host spin defects with energy levels that enable transitions in the microwave photon energy regime. Since SiC can host a variety of different spin defects which reveal EPR signatures, these defects are perfect candidates for a potential maser. One of the most investigated spin defects is the silicon vacancy V2 in the polytype 4H SiC. Especially, the spin polarization of this system under optical excitation is substantial for a gain medium. The pumping scheme is shown in Fig. 1b. The system possesses a quartet ground spin state ($S = 3/2$), which can be optically excited by photons in the near infrared. [18,20]

2) Optical pumping: Since the excited state relaxes spin-dependently, we end in a spin polarization of the ground state, with the energetically lower laying $J = 1/2$ manifold being preferred. [21] Similar to a laser, a maser also requires a population inversion. To achieve this inverted population difference, we shift the higher populated energy level above the less populated energy level via applying an external magnetic field of $\gg$2.5 mT (corresponding to 70 MHz). This overcomes the ground state level anti-crossing (GSLA), where the order of energy levels switches due to the Zeeman effect. [22] If now an incoming microwave photon possesses the same energy as the Zeeman-split transition energy, it leads to an avalanche of coherent microwave photons. [14] The initial photon may originate either from an external microwave signal, when the maser operates as a low-noise amplifier, or from spontaneous emission, which seeds stimulated emission when the maser functions as a coherent microwave source. To achieve a sufficient coupling of the microwaves to the spin system, a resonator with a high quality is required.

3) High-Q resonator: The resonator for the here-reported SiC maser consists of two parts: a cylindrical metallic copper resonator and a dielectric sapphire core in the center which contains the SiC gain material. The sapphire ring concentrates the microwaves, reducing the ohmic losses in the copper wall and thus increases the Q-factor, which is the figure of merit for the resonator's quality. The magnetic field components of the microwaves are illustrated in purple in Fig. 1a. A coupling antenna outcouples the created microwaves into a standard SMA cable to enable characterization of the microwave output and finally use the maser as a coherent microwave source. A detailed analysis of the resonator is given in the Supplementary Figs. 7-12.

## SiC-based maser

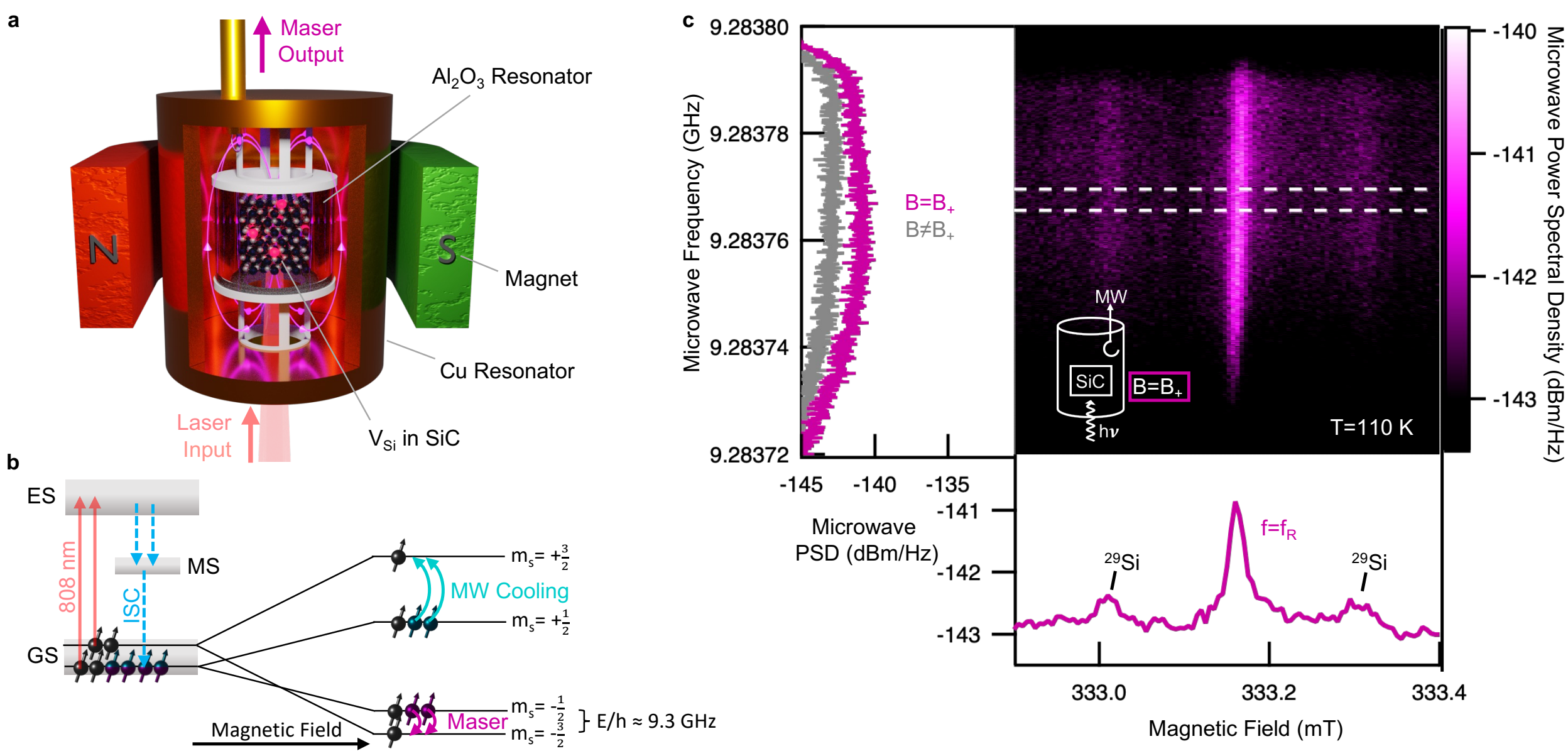

**Fig. 1. Silicon carbide-based maser. a** Basic principle: maser output is achieved by optical pumping of the gain material (SiC) in a high-Q resonator inside an external magnetic field. **b** Pumping scheme of the gain material: a population inversion is created by an external magnetic field which can be used for the maser process (pink). The second spin polarized transition (cyan) can be used for a microwave refrigerator. **c** Maser output measurement while sweeping the magnetic field at $T = 110$ K: the color map reveals a broad emission peak displayed in the frequency domain (left plot) and over the magnetic field (bottom plot). The microwave power plotted vs. the magnetic field is the average maser output power in between the two dashed lines. A narrow peak is observed at the resonant transition $B = B_+$ with a linewidth of 0.021 mT and two satellite peaks according to hyperfine interaction with $^{29}Si$ isotopes. The inset illustrates the setup configuration. A detailed description can be found in Supplementary Fig. 4.

In order to demonstrate a SiC-based maser, microwave amplifier and refrigerator in this work, the standard measurement method is the detection of the microwave intensity in a specific frequency band with a spectrum analyzer attached to the end of the SMA cable of the coupling antenna (see Supplementary Figs. 1-6 for detailed setup description). The external magnetic field is tuned by using magnetic field coils within the range of the expected emissive/absorptive transition. In principle an arbitrary magnetic field can be used above the ground state level anti-crossing (GSLA). However, the corresponding frequency has to fit to the resonance frequency of the resonator which is in the case 9.3 GHz corresponding to a resonant magnetic field of about $B_+ \approx 333$ mT. The resonator is incorporated in a cryostat to perform measurements at lower temperatures ($T = 110$ K). In this environment the population inversion is easier to achieve due to the longer spin-lattice relaxation times. [23] Furthermore, the Q-factor is higher since the ohmic losses are reduced in the copper walls. The power spectral density (PSD) of the noise level due to thermal microwave photons is around -143 dBm/Hz. Fig. 1c shows the frequency resolved output power of this configuration for a swept magnetic field around the resonant condition ($B_+ = 333.19$ mT) and a sufficient high pump power ($P_{\mathrm{Laser}} = 450$ mW) of an 808 nm excitation. The intensity of the color map represents the microwave output power which reveals a pattern with three narrow features: one pronounced signal in the center and two outer peaks. To study it more in detail the bottom figure reveals the average maser power within the two dashed white lines. We can now clearly extract a narrow peak with a linewidth of 0.021 mT (corresponding to 588 kHz) and a total amplitude of -141 dBm/Hz. This small signal is attributed to the silicon vacancy since it arises at the correct magnetic field and shows two additional satellite peaks. These are related to the $^{29}$Si isotopes surrounding the silicon vacancies with a natural abundance of 4.7%, leading to a shifted resonance condition due to hyperfine splitting (150 µT). With this measurement, we demonstrate the first SiC-based maser in a CW mode. However, as we can see for the slice in the frequency domain (left subfigure of Fig. 1c) the signal stands out just marginally from the thermal background. To enhance the output power and realize room temperature masing, we address critical parameters such as optical pumping and the Q-factor of the resonator.

### Q-boosted maser

At low temperatures, the generation of microwave photons $|a|^2$ can be quantified by solving the Langevin equation, which can be adapted for describing a maser output as derived in [24]:

$$|a|^2 = \frac{\omega - \gamma_{\mathrm{eg}}}{2\frac{\omega_{\mathrm{c}}}{Q}} N - \frac{\omega + \gamma_{\mathrm{eg}}}{2\frac{\omega_{\mathrm{c}}}{Q}} S_z \qquad [1]$$

Here, the pump rate $\omega$ is given by the effective optical pump power $P_{Laser} = \omega\frac{hc}{\lambda}$, $\gamma_{\mathrm{eg}} = T_1^{-1}$ is the spin lattice relaxation rate, $\omega_{\mathrm{c}}$ and $Q$ are the frequency and quality factor of the resonator, respectively. $N$ is the number of participating spins and $S_z = \frac{\kappa_{\mathrm{S}}\kappa_{\mathrm{c}}}{4g^2}$ the number of polarized spins mainly given by the coupling constant $g$ and the collective decay rate $\kappa_{\mathrm{S}} \approx \frac{2}{T_2^*}$ limited by the coherence time $T_2^*$. In order to realize a working maser, a microwave output is required, thus $|a|^2 > 0$ defines the maser threshold. Inserted into equation [1] the following expression can be derived as limit for the necessary Q-factor:

$$Q > \frac{\omega + \gamma_{\mathrm{eg}}}{\omega - \gamma_{\mathrm{eg}}} \frac{\kappa_{\mathrm{S}}\omega_{\mathrm{c}}}{4Ng^2} \qquad [2]$$

In Fig. 2a, we show the expected microwave power according to equation [1] for our SiC maser for a broad range of different Q-factors and optical pump powers. Two thresholds are relevant: the pump threshold $\omega > \gamma_{\mathrm{eg}}$ (vertical grey dashed line) and the maser threshold according to equation [2] (black dashed curve). The pump threshold is mainly given by the $T_1$ time and can be exceeded by increasing the pump power. The maser threshold, however, strongly depends on the coupling and the coherence time $T_2^*$ of the system and can be reached by increasing the quality factor of the resonator.

First, we use the optical pump power threshold $P_{th}$ to determine the efficiency of creating a population inversion. To keep up with the spin-lattice relaxation of a single spin, we need $1s \cdot \gamma_{\mathrm{eg}}$ photons per second, each with an energy of $\frac{hc}{\lambda}$. This value is scaled by $N$ to determine the total power required for the entire ensemble, and subsequently divided by the input power used in the experimental setup to achieve this condition:

$$\eta = \frac{N\,\gamma_{\mathrm{eg}}\frac{hc}{\lambda}}{P_{th}} \qquad [3]$$

The extracted pump efficiency of approximately 0.4% quantifies how effectively an incident optical photon contributes to microwave photon generation. This limited efficiency primarily arises from optical losses in the excitation path and incomplete photon absorption, including reflection at the SiC surface and transmission through the sample.

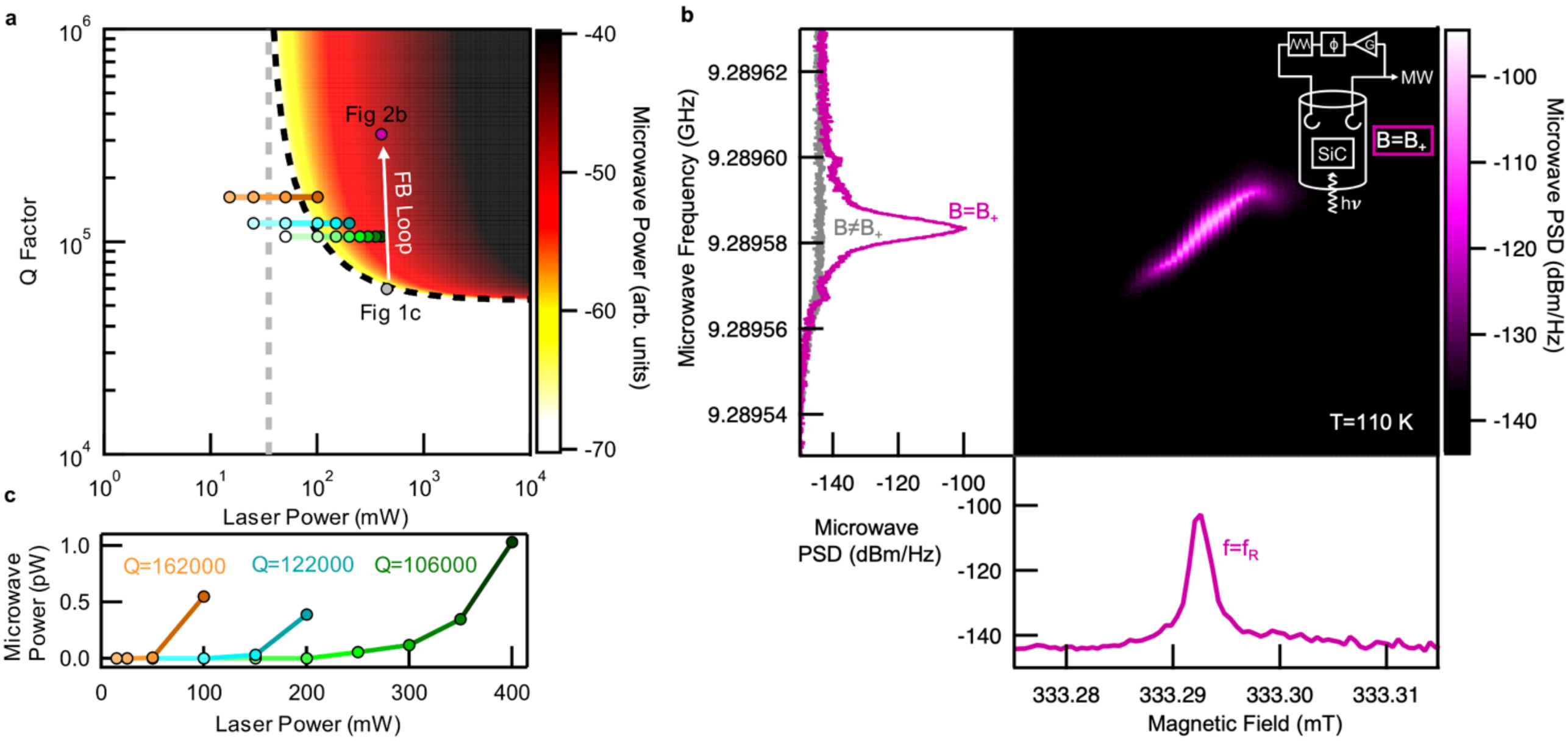

**Fig. 2. Q-boosted maser at $T = 110$ K. a** Theoretical model of a SiC-based maser: the masing regime is indicated by the black dashed curve while the grey dashed line shows the pump threshold for achieving a population inversion. **b** Maser output of a feedback-loop (FB) driven maser at 110 K: the left sub-plot displays a sharp peak in the frequency domain while the bottom graph shows the in-resonance cross-section over the magnetic field. The inset illustrates the setup configuration. A detailed description can be found in Supplementary Fig. 5. **c** Observed maximum maser output for a large range of optical pumping power and Q-factors. All measurements are inserted in the theoretical model in **a**.

Inserting the whole parameter set (see Supplementary Figs. 12-15) into [2] we can explain the very low maser power observed in Fig. 1c with a maser operating directly at the maser threshold as shown in Fig. 2a. The measurement of Fig. 1c and the following measurements are inserted as circles into the model. We now aim to enhance the SiC maser by further entering the masing regime. This improvement can be achieved either by increasing the coherence time or by changing the resonator properties. The first can be realized by isotopical purification of $^{28}$SiC [25,26] (which would in addition nearly double the effective number of spins, as only the main transition of $V_{Si}$ associated with $^{28}$Si is used). A more continuous method of probing the maser performance involves changing the quality of the resonator in use. A very powerful tool is the implementation of a feedback-loop enabling an artificial enhancement of the Q-factor (see Supplementary Fig. 5 for details). [27,28] In general, the Q-factor is defined by the energy losses per cycle. However, if the resonator is fed with its own microwave output, the energy losses can be reduced and the Q-factor is enhanced artificially. This is achieved by an outcoupling of the microwaves with the same antenna as in the previous measurement, but half of the signal (-3 dB) is amplified and inserted phase-corrected back into the resonator. Importantly, the use of a low-phase-noise amplifier (LNA) in the feedback loop is crucial to preserve spectral purity; otherwise, phase noise introduced by the loop could degrade the maser's performance. However, as shown later, the noise performance of the maser is significantly worse than that of the used LNA due to the maser's magnetic field sensitivity. Therefore, the additional amplifier noise introduced by the feedback loop is not relevant to the current configuration. With this technique, we are able to boost the Q-factor of measurement Fig. 1c by a factor of 5 which is illustrated in Fig. 2a (grey and purple circles).

The feedback-loop driven maser measurement of the frequency resolved output power versus the swept magnetic field is depicted in Fig. 2b. We choose a narrower magnetic field range in order to study the shape of the microwave peak in magnetic field direction. The color map reveals an S-shape pattern which is due to the overlap of the Lorentzian peak contribution in f-direction and a Lorentzian peak in B-direction due to the natural line width of the EPR transition. [14] The output power approaches almost -100 dBm/Hz making it already usable for practical applications. A cross section through the center of the S-shape is depicted in the bottom sub-plot of Fig. 2b. The linewidth 0.001 mT of the microwave output is much narrower than in the previous measurement of Fig. 1c (factor of 21). We performed several feedback-loop driven maser measurements (see Supplementary Fig. 19) to study the influence of Q-factor and microwave power which are depicted in Fig. 2c. We can observe that the maser output strongly depends on the Q-factor which coincides with the simulation. For higher Q-factors, a lower laser power is required to reach the maser threshold. The measured values are also inserted into the simulated masing regime. However, the output power is still lower than the expected ideal output power of the simulation which is mainly owed to coupling losses of the antenna but also losses between the many connectors of the SMA components of the setup (see Supplementary Information for setup details). Nevertheless, we can present a drastic improvement of the signal quality and strength by an artificial enhancement of the Q-factor which reveals the large potential of this SiC-based maser.

### SiC-based maser at room temperature

All results presented so far were obtained at cryogenic temperatures ($T = 110$ K). To enable masing at room temperature, we first investigate the temperature dependence of the characteristic pump power $P_0$ which is key to understanding the pump requirements. Fischer et al. [29] proposed that $P_0$ is proportional to the spin-lattice relaxation rate $T_1^{-1}$. To verify whether this relationship holds for continuous optical pumping across a broad temperature range, we performed EPR measurements at various laser powers and temperatures to extract $P_0$ (see Supplementary Fig. 14). The results are shown in Fig. 3a.

We find that $P_0$ closely follows the known temperature dependence of $T_1^{-1}$, as reported by Simin et al. [23], supporting the assumption that $T_1^{-1}$ is the dominant relaxation mechanism in the previously discussed model. The pump power dependence at low (blue) and high (red) temperatures shows slight deviations from the expected behavior. At low temperatures, the fit begins to transition into the constant $A_0$ regime. While closer to ambient temperatures, we observe a power law exponent slightly lower than the $T^5$ dependence, likely due to an accelerated $T_1^{-1}$ rate under continuous optical pumping. Overall, these results suggest that the effective $T_1$ time is overestimated at lower temperatures, leading to a reduced pump power requirement at higher temperatures, which is beneficial for reaching the maser at room-temperature operation.

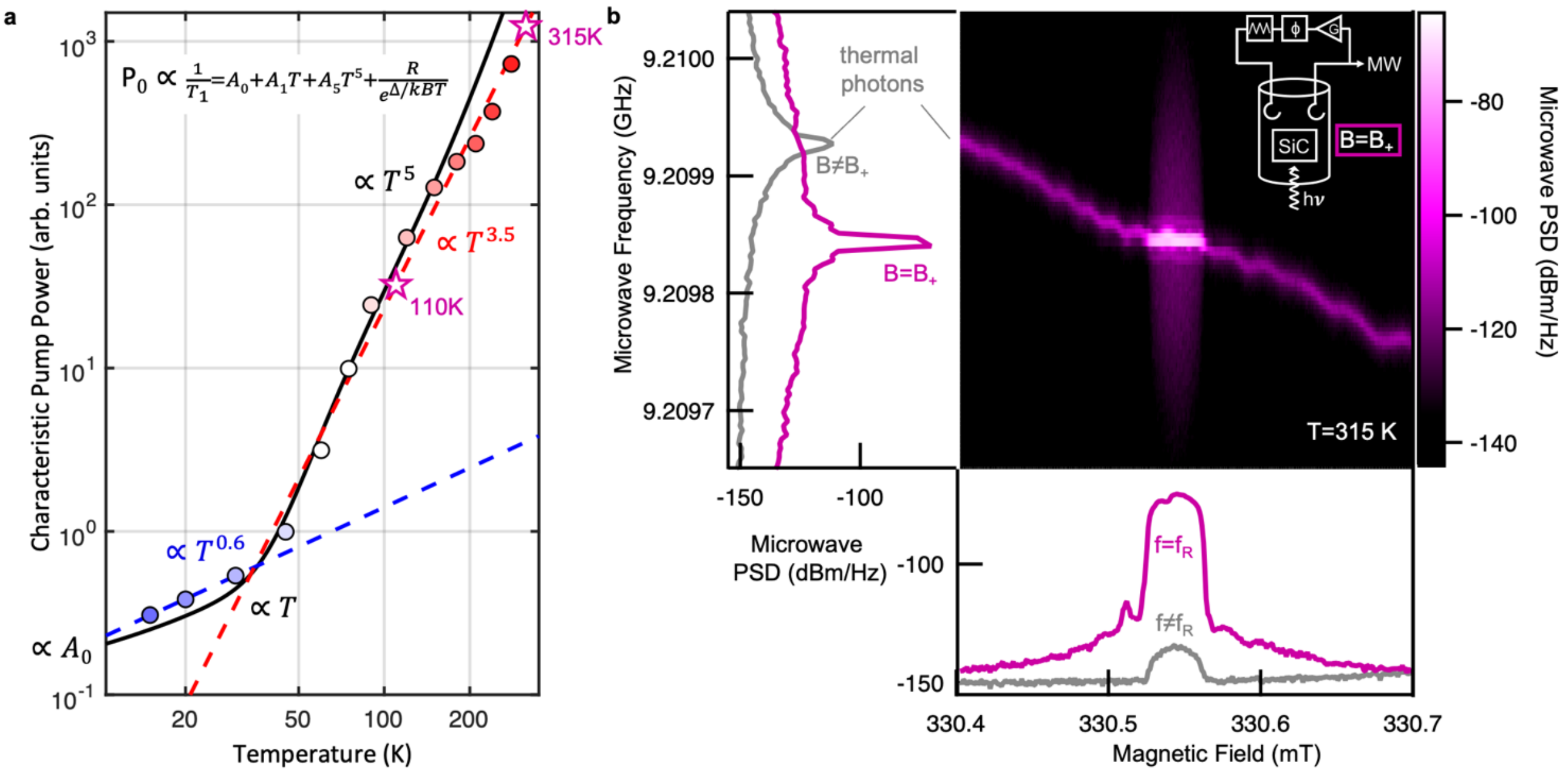

**Fig. 3. Towards a room-temperature maser. a** Temperature dependence of characteristic pump power $P_0$ using EPR. Qualitatively, $P_0$ directly follows the $T_1$ relaxation rate as reported by [23]. Quantitatively, we observe reduced exponents in the high (red) and low (blue) temperature regime due to accelerated $T_1$ relaxation under continuous optical pumping. Demonstrated maser operation at $T$ = 110 K and 315 K is indicated by the star symbols. **b** Q-boosted maser at $T$ = 315 K: The color map reveals a high output power of -80 dBm/Hz (central bright spot) at $f = f_R$ and $B = B_+$. The purple trace is given by the amplified thermal photons inside the cavity (see text for details). A cross section at a non-resonant magnetic field $B \neq B_+$ (grey) shows an enhancement effect of thermal microwave photons.

To reach the maser threshold we increase the gain of the feed-back loop, which results in the first maser output at room-temperature when pumped sufficiently with 617 mW (see Fig. 3b). At this temperature (315 K because of optical pumping), the resonance condition occurs at lower magnetic fields due to the lower resonance frequency of the thermally expanding resonator dimensions (see Supplementary Fig. 11 for temperature dependence and Supplementary Figs. 17-20 for Q-factor and pumping power dependence). The observed signal contains two features: first, a very intense microwave output at the resonant conditions ($f = f_R$ and $B = B_+$) and second, a continuous trace with variable frequency. The first signal (the actual maser output) reveals a high microwave output of -80 dBm for a bandwidth of 1 Hz. This is related to the very sensitive feedback loop which has to be set to a high amplifying mode in order to reach the maser threshold at room temperature. This amplification mode is visible in a cross-section, as depicted in the bottom sub-plot of Fig. 3b. A substantial increase of the broadening can be observed in comparison to the previous measurements. The reason is, that the signal is already reaching the maser regime with the Q-boosted resonator for magnetic fields at the edges of the magnetic resonance condition. The maser output with frequency resolution, however, is narrower as displayed on the left of Fig. 3b. This linewidth can be considered as the actual linewidth of the maser, representing the frequency domain used in practical applications. The broad plateau with its linewidth of 0.04 mT is in fact a large advantage since it demonstrates the stability of the maser over magnetic field fluctuations. If this peak were narrower, the maser output would collapse for a small magnetic field deviation which limits the application of the proposed maser device. The most critical component we observed is the stability of the resonator frequency of our device. If the frequency shifts due to temperature fluctuations/drifts, the whole maser frequency is shifting accordingly. This resonance frequency change can be observed by analyzing the continuous random trace, which originates from thermal photons within the resonator. These photons are slightly amplified by the highly sensitive feedback loop and fluctuate randomly over time, independent of the magnetic field. As thermal fluctuations affect the resonator's physical dimensions, they cause small shifts in its resonance frequency, e.g., the observed 200 kHz drift at 9.2 GHz corresponds to a ≈0.002% deviation. Although this signal is several orders of magnitude weaker than the maser emission, it does not interfere with operation and can be used to track the resonance frequency without performing Q-factor measurements. Therefore, the random trace provides valuable insight into frequency stability, highlighting a critical aspect of maser performance. In future applications, this signal could be fed into a PID control loop to actively stabilize the resonator (such as by adjusting the cavity height with a piezoelectric actuator) ensuring a stable and reliable maser output frequency. The overall maser performance can further be improved by using a resonator with higher Q by treating the surface with electro polishing and by enhancing the coherence time of the sample via isotopic purification to lower the maser threshold by orders of magnitude. [26] After this initial demonstration of the first SiC maser, we want to discuss potential applications.

**SiC-based maser applications**

1) Amplifier: Besides the application of a maser as a coherent microwave source, the SiC maser can be used as a low noise amplifier of weak microwave signals. The configuration is identical to the setup of the feedback-loop, with the exception that an external signal is fed into the resonator instead of its own microwave output (see inset in Fig. 4a and Supplementary Fig. 4 for details). We analyzed this behavior by inserting weak microwave signals with one antenna and measuring the output by the second antenna while sweeping the magnetic field. The result is depicted in Fig. 4a. The horizontal line is the inserted constant microwave signal (which is within the linewidth of the resonator). While sweeping the magnetic field, we can observe an amplification of the signal at the resonant magnetic field. A cross section is shown in the bottom sub-plot of Fig. 4a. We choose the magnetic field sweep broad enough to also study the influence of the $^{29}$Si hyperfine peaks. The optically-pumped gain medium amplifies the signal by 7 dB (1.8 dB for hyperfine peaks) which makes it ideal for a low-noise preamplifier. No broadening of the signal can be observed, as depicted in the left sub-plot of Fig. 4a ($B = B_+$: amplifier on, $B \neq B_+$: amplifier off).

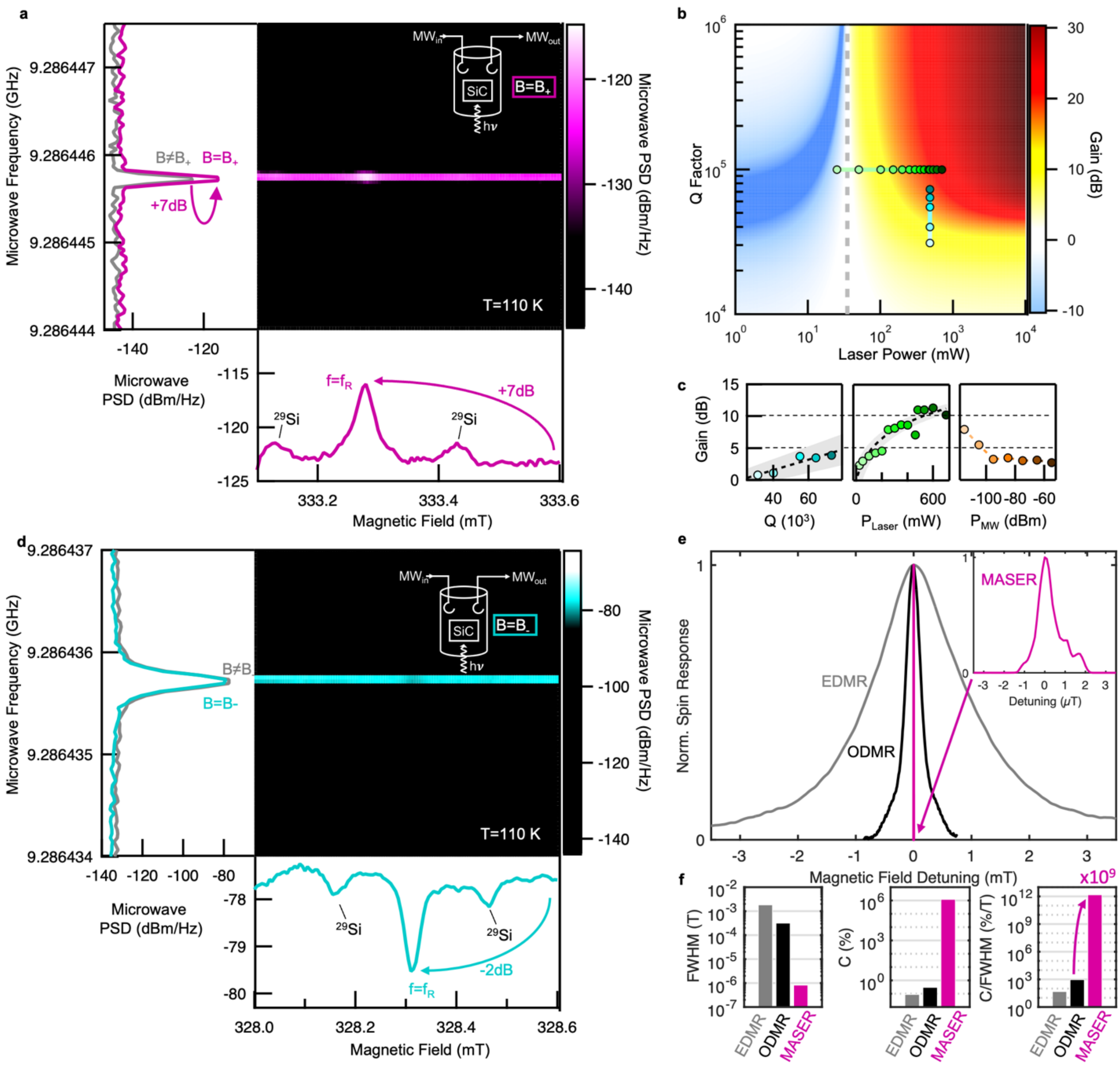

**Fig. 4. SiC-based amplifier / refrigerator using the maser transition at $T = 110$ K. a** Color map of the measured microwaves: a narrow purple line is given by the inserted microwave signal. As soon as the system is in resonance, the signal is amplified by the maser (bright white spot at $B = B_{+}$). The cross sections in the frequency domain and over the magnetic field are displayed in the left and bottom sub-plots, respectively. The insets in **a, d** illustrate the setup configuration. A detailed description can be found in Supplementary Fig. 6. **b** Theoretical model of the amplifier: the color represents the expected amplification/attenuation. **c** Measured gain depending on Q-factor, pump power ($P_{\mathrm{Laser}}$) and signal input power ($P_{\mathrm{MW}}$), respectively (input power of -70 dBm, unless stated otherwise). **d** SiC-based refrigerator with color map of the measured microwaves: a narrow cyan line is given by the inserted microwave signal. As soon as the system is in resonance the signal is attenuated due to induced spin flips of the $V_{Si}$. The cross sections over frequency and magnetic field are displayed in the left and bottom sub-plots, respectively. **e** Magnetometer performance of the maser (slice from Fig. 3b) in comparison to electrical and optical readout techniques (EDMR and ODMR).[31,32] **f** Linewidth ($\mathrm{FWHM}$), readout contrast ($C$), and ratio $C/\mathrm{FWHM}$ of the difference system. Maser offers a readout enhancement by nine orders of magnitude compared to existing systems using the same sample type.

To reveal the full potential of this preamplifier, we simulated the gain for different Q-factors and different pump powers which is shown in Fig. 4b. In principle, the amplifier can amplify weak signals by several orders of magnitude. This theoretical model is verified by further measurements with varied Q-factors (shown in blue) and variable pump power (shown in green). The corresponding measurements are displayed in Fig. 4c. The amplification can be enhanced by increasing the Q-factor as well as by increasing the optical pump power. Thus, we observe the highest amplification of the signal by 11 dB for a laser power of 500 mW. In general, we can also observe and simulate that the amplification is higher for weaker signals which is shown with the orange circles in Fig. 4c (see Supplementary Fig. 16 for corresponding simulations). The noise figure of this amplifier ranges from 1 dB to 3.5 dB with a noise temperature well below the device temperature (see Supplementary Fig. 21). It still operates well above the quantum limit (0.5 K), in contrast to the diamond-based amplifier reported by [30]. Further advances in SiC material engineering are expected to bring performance closer to this limit. For example, through the use of isotopically purified SiC, which offers up to a 50-fold increase in coherence time. [26] An extended coherence time, in turn, enhances the amplifier's gain and drives its noise temperature toward the quantum limit. Such developments would establish SiC as a highly competitive platform, combining a quantum limited performance with the inherent advantages of a semiconductor.

2) Microwave refrigerator: Besides using SiC as a coherent microwave source and a low-noise amplifier, this system can also be utilized in reverse as an optically-pumped microwave absorber. This leads to an effective cooling of the microwave mode temperature and can be used as a refrigerator, as described in the following. To demonstrate an optically-pumped microwave absorber, we use the same setup as for the amplifier (see Fig. 4a) but a slightly shifted magnetic field value ($B = B_-$). Since silicon vacancies in SiC possess a $S = 3/2$ spin system, there are three possible ($\Delta m_s = \pm 1$) microwave transitions: $B_-, B_0, B_+$. While the latter transition is emissive and can be used for an amplification of microwaves (as described and used before), the two others are absorptive transitions. The center transition $B_0$ carries a population difference given by the Boltzmann statistic. Therefore, the spin temperature $T_{\mathrm{spin}}$ of this transition is always given by the environmental temperature $T_0$. [19,29] The $B_-$ transition, however, can be optically spin polarized to the lower level and can thus be used to absorb microwave photons, thereby reducing the number of photons in the cavity.

To study this effect and the potential of SiC as a mode cooling device, we insert a microwave signal into the cavity by one antenna and measure the outgoing microwaves after interacting with the spin system with the second antenna. The corresponding color map for a magnetic field sweep over the absorptive transition ($B_-$) is displayed in Fig. 4d. Again, the cross sections through the resonance position are displayed in the left and bottom sub-plots, respectively. The attenuation of the signal over the frequency can in fact be barely registered and seems to be in the range of noise. Whether this effect originates from the defect can be observed more clearly if the cross-section is set along the incident frequency and the signal is plotted over the magnetic field. In the bottom sub-plot the silicon vacancy transition with its hyperfine peaks can clearly be observed, resulting in an attenuation of -2 dB. This low attenuation already corresponds to a temperature reduction of $\Delta T = 40$ K, well below the device temperature.

In this context, we want to highlight that a unique feature of the silicon vacancy is the very small ZFS. This results in a small required magnetic field offset of just ±2.5 mT to switch between the refrigerator and the amplifier, which might enable interesting future applications for this kind of novel microwave device. In contrast, NV diamond is not applicable for an easily switchable amplifier/refrigerator device since way larger field offsets of ±102.5 mT are required. Looking further, the V1 defect shows even more promise for this application than the here-studied V2 defect due to its even smaller ZFS. Therefore, a similar maser device can be realized using the V1 defect with the advantage that switching between emission/absorption requires only ±163 μT as shown in Supplementary Fig. 22.

3) Maser magnetometer: In all previously demonstrated applications, we observed instability in the maser output not only due to resonator fluctuations but also because of magnetic field noise. Therefore, unlike the hydrogen maser, this SiC maser (or spin-defect masers in general) is not suitable as an oscillator for timing applications. However, this limitation reveals a new opportunity in quantum metrology. Since small variations in the magnetic field directly translate to strong frequency or output power shifts, this maser can instead be used for magnetometry.

While the use of spin resonance frequencies for magnetometry is well established, leveraging the maser output offers distinct advantages. To demonstrate its potential, we compare this new maser-based readout with conventional readouts from the same material system (SiC) at ambient temperatures. The most common readout techniques for spin-defect-based magnetometry are optically detected magnetic resonance (ODMR) and electrically detected magnetic resonance (EDMR). One key parameter defining the spin response is the system's linewidth ($\mathrm{FWHM}$). A comparison of representative ODMR and EDMR spectra is shown in Fig. 4e (see [31,32] for details). The EDMR signal (gray trace) exhibits a typically broad response, while ODMR (black trace) shows a sharper response when detuned from resonance. The room-temperature maser, however, is another two orders of magnitude narrower than ODMR and only resolvable when zooming into the magnetic field axis (see inset). A comparison of linewidths is presented in the first subpanel of Fig. 4f. While the ODMR linewidth is fundamentally limited by the system's coherence time, the maser linewidth is significantly narrower due to superradiance and is constrained only by the Schawlow–Townes limit, which is indirectly related to coherence time. Another critical parameter of spin response is the contrast, $C$, which quantifies the relative signal change between on-resonance and off-resonance conditions. The ODMR contrast in SiC (defined as $C = \Delta\mathrm{PL}/\mathrm{PL}$ with $\mathrm{PL}$ being the measured photoluminescence in off condition) is typically below 1%, whereas other systems, such as NV centers in diamond or boron vacancies in hBN, achieve values in the 10 – 50% range. [33] EDMR performs even worse; the spin-dependent recombination current $\Delta I$ is small compared to the device current $I$, resulting in a contrast $C = \Delta I/I$ of less than 0.1%. In both ODMR and EDMR, the signal source (photoluminescence or current) remains active regardless of resonance, with only slight changes indicating resonance. The maser, by contrast, operates exclusively under resonance conditions and switches off entirely when off-resonance. In the off-state, the residual signal consists solely of thermal photons, previously observed as random noise in the measurement (Fig. 3b). Even when using the maximum off-resonance signal level as a baseline, the resulting contrast is $10^6$%, offering a readout performance that is several orders of magnitude superior to other techniques. Notably, the ratio of contrast to linewidth ($C/\mathrm{FWHM}$) serves as a figure of merit. In ODMR, high contrast is typically achieved using strong microwave power, which causes power broadening and hence a broader resonance peak. Conversely, achieving a $T_2^*$-limited linewidth often comes at the expense of reduced contrast. As a result, a trade-off must be made to optimize this ratio. The maser, however, offers both high contrast and narrow linewidth simultaneously, resulting in a performance improvement of up to nine orders of magnitude within the same spin system. The sensitivity of the maser-based magnetometer is ultimately limited by laser noise from the excitation source. Accounting for this contribution yields a magnetic-field sensitivity of 20 pT/√Hz (see Supplementary Fig. 23 for details). This value surpasses existing SiC readout techniques by orders of magnitude and lies within the theoretically predicted performance range for comparable systems. [24] An additional noise source contributing to an effective broadening of the linewidth arises from fluctuations in the bias magnetic-field coils. Maser implementations that operate without a bias field, such as pentacene in *p*-terphenyl, achieve sensitivities of 42 pT/√Hz , [28] but are limited to pulsed operation. An ideal platform would combine material robustness with a pumping scheme that enables population inversion without the need for a bias field.

In summary, our study marks a significant milestone by introducing the first continuous-wave semiconductor maser based on silicon vacancies in silicon carbide. Employing a feedback loop, we successfully enhance the Q-factor, enabling precise control of critical parameters necessary for a robust and coherent microwave source, even beyond room temperature. Through detailed simulations, we highlight the broad potential of SiC-based maser technology in applications such as signal amplification, mode cooling, and magnetometry. Notably, in the latter, our approach offers a remarkable nine orders of magnitude improvement over alternative electrical and optical readout techniques. SiC's established role in the semiconductor industry and its compatibility with large-scale manufacturing underscore its promise as a platform for scalable, room-temperature masers. Furthermore, the spin defects employed here can be polarized not only optically but also electrically, opening a pathway toward the development of electrically driven, on-chip masers. [34] Just as the transition from optical lasers to compact laser diodes transformed everyday technologies, our work lays the foundation for maser diodes, potentially initiating a new field of practical maser-based applications across science, engineering, and industry.

## DATA AVAILABILITY

The data generated in this study are provided in the Supplementary Information / Source Data file.

## FUNDING

This research was funded by the European Research Council (ERC) (Grant agreement No. 101055454) and the German Research Foundation DFG (DFG-DY 18/13-1). A.G., A.S. acknowledge financial support from the DFG through the Würzburg-Dresden Cluster of Excellence on Complexity, Topology and Dynamics in Quantum Matter - ctd.qmat (EXC 2147, project-id 39085490). The magnetometer analysis was carried out at the Jet Propulsion Laboratory, California Institute of Technology, under contract with the National Aeronautics and Space Administration (contract 80NM0018D0004) and funded by the Planetary Instrument Concepts for the Advancement of Solar System Observations (PICASSO) program.

## ACKNOWLEDGEMENTS

We thank Paul Konrad for useful discussions.

## AUTHOR CONTRIBUTIONS

A.G., M.W. and A.S. designed and realized the home-made spectrometer. A.G. and M.W. performed theoretical calculations of the relevant parameters affecting the maser threshold and performed all measurements. V.B. designed the microwave resonator which was characterized by A.G.. E.E. contributed with additional temperature dependent EPR measurements. A.G., V.D. and A.S. wrote the manuscript. A.S. conceived the idea and supervised the project together with A.G.. All authors discussed the results and commented on the manuscript.

## COMPETING INTERESTS

The authors declare no competing interests.

# Supplementary Information for:

# Semiconductor Room-Temperature Maser

Andreas Gottscholl[1,2,a)], Maximilian Wagenhöfer[1], Valentin Baianov[1], Emilian Eisermann[1], Vladimir Dyakonov[1], Andreas Sperlich[1,b)]

[1]Experimental Physics 6 and Würzburg-Dresden Cluster of Excellence ctd.qmat, Julius-Maximilians-Universität Würzburg, 97074 Würzburg, Germany

[2]NASA Jet Propulsion Laboratory, California Institute of Technology, Pasadena, California, USA

[a)] Electronic mail: gottscholl.andreas@gmail.com

[b)] Electronic mail: andreas.sperlich@uni-wuerzburg.de

## Setup Details

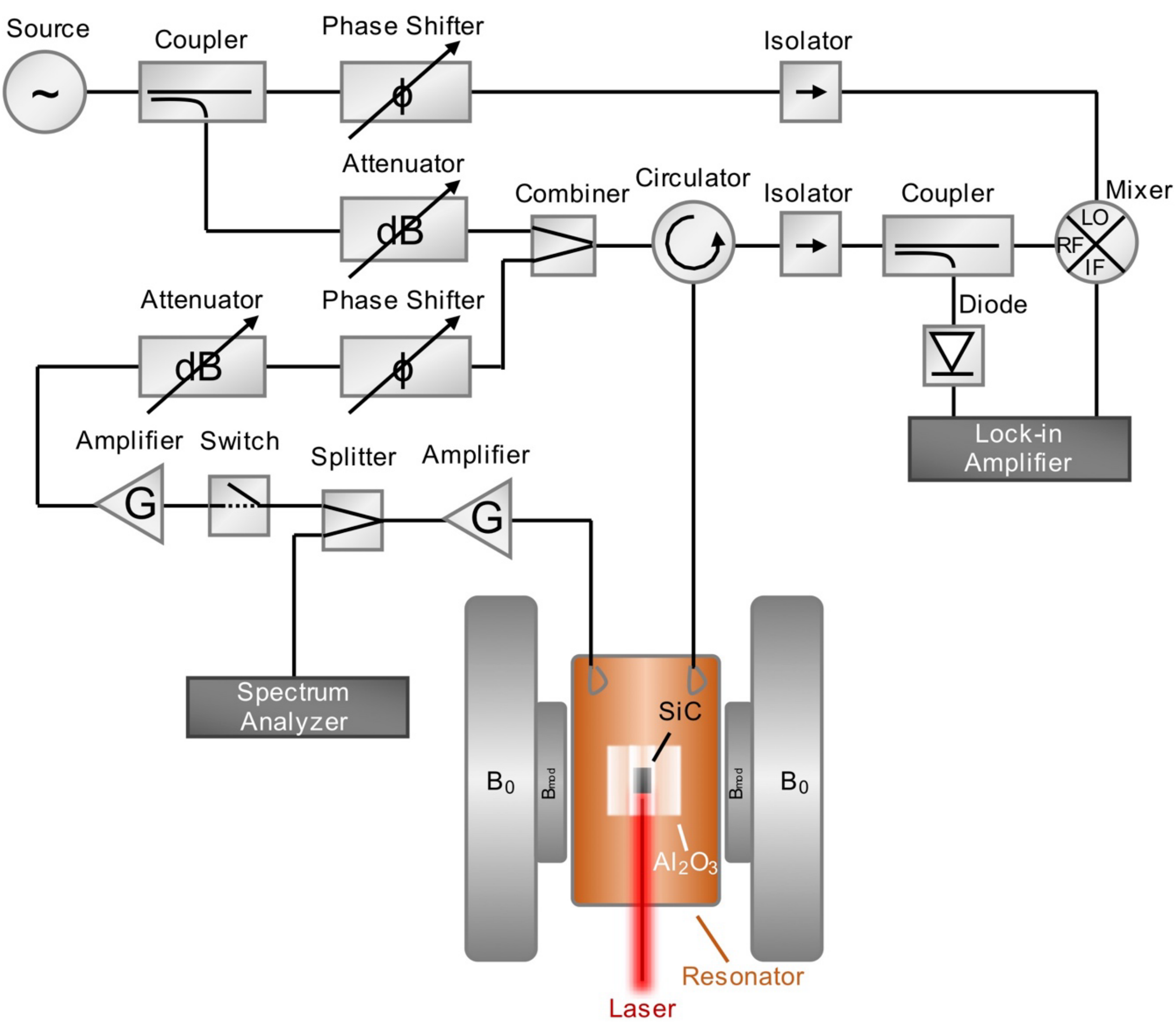

**Supplementary Figure 1. Self-built setup for Q-factor, electron paramagnetic resonance, (Q-boosted) maser, amplifier and refrigerator measurements.**

The main results of this work were all recorded with one single self-built setup. Since the here-reported microwave bridge enables six different modes of operation, we will focus on the different signal paths separately for each measurement in the next sections. The whole setup is illustrated in Supplementary Fig. 1. As a microwave source we used an *Anritsu MG 3694C*. Most of the components are commercially available: As coupler we used a -10 dB *Mini-Circuits ZUDC10-02183-S*+ (left) and a -20 dB *Radiall R433724700* (right). The isolators *Aerotek I2E1L1FF* are integrated to block unwanted microwave reflections. For combiner/splitter we used *Mini-Circuits ZFRSC-183-S+,* while the detection is performed with a detector diode *Advanced Control Components ACTP-1504P*, a mixer *Marki M10412LA*, a lock-in amplifier *Signal Recovery 7230* and a spectrum analyzer *WSA 5000* (see the following section for detailed measurement explanations). To compensate microwave losses, we used a *Miteq AFS4.08001200-10.200* LNA (right) and a *Narda N62448-243* (left). A microwave switch *Mini-Circuits MSP2TA-18-12*+ as well as two modified Rotary Vane attenuators with added servo motors are controlled via an *Arduino*. For phase corrections we used two analog phase shifters from an old *Bruker ER047* microwave bridge, which are controlled via the DAC of the lock-in amplifier. For optical pumping we used an 808 nm-laser *K808FANFA-15.00W* from *BWT* which provides a variable power of 0 – 15000 mW. The resonator is placed in a cryostat with an optical window on the side. For low temperatures, the cryostat walls are filled with liquid nitrogen, and the cryostat volume is maintained in a helium atmosphere to efficiently transfer the temperature to the resonator and the sample. The temperature is monitored using a *Cernox* temperature sensor attached to the exterior of the resonator. The microwaves are coupled in and out of the resonator via loop antennas.

**Q-factor measurement**

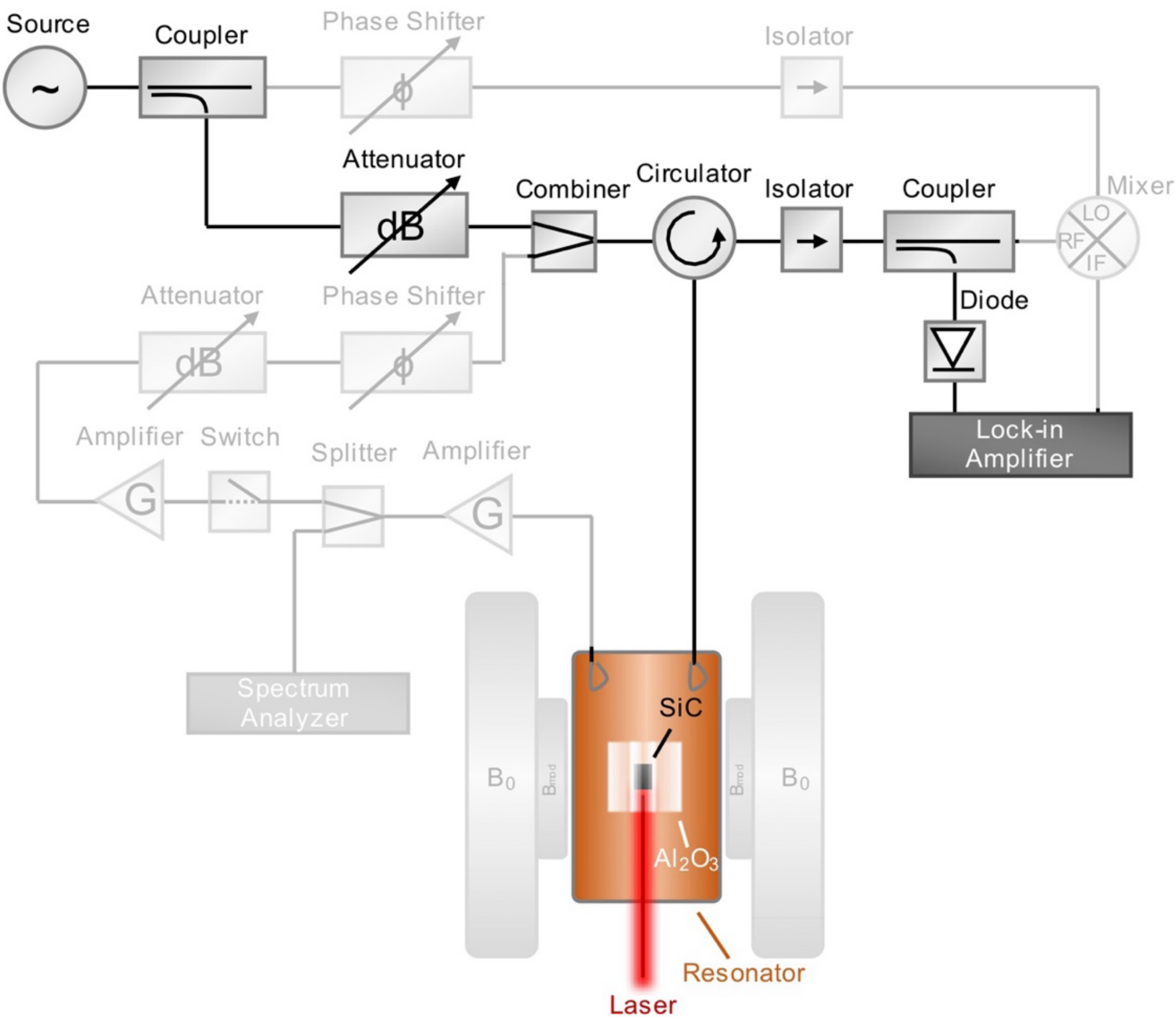

**Supplementary Figure 2. Setup for Q-factor measurements.** The microwave frequency is swept while the output is on/off-modulated. The reflected microwave power is measured via a detector diode and a lock-in amplifier.

The basis for all measurements reported in this paper is the determination of the Q-factor of the resonator and in particular its resonance frequency. This is accomplished by sweeping the frequency of an inserted microwave probe while measuring the reflected microwave power. Via a circulator the reflected microwaves are separated from the inserted microwaves and detected with a detector diode. Since the measurement provides a very high signal-to-noise ratio only a very small amount of the signal is required. Thus, we use a coupler with -20 dB which transmits most of the signal for the EPR measurement (see next section). The detection of the small signal is realized with a lock-in amplifier. We choose an on/off modulation of the microwaves in order to observe the reflection directly, which results in a quantitative analysis of the Lorentzian dip providing information about under/over-coupling or critical coupling of the resonator (see Supplementary Fig. 9). This calibration routine for the maser/amplifier/refrigerator is performed with an active laser output since a small heating input leads to a drift of the resonance frequency. Thus, we start with further measurements as soon as the Q-factor measurements reveals a stable resonance frequency.

## Electron Paramagnetic Resonance

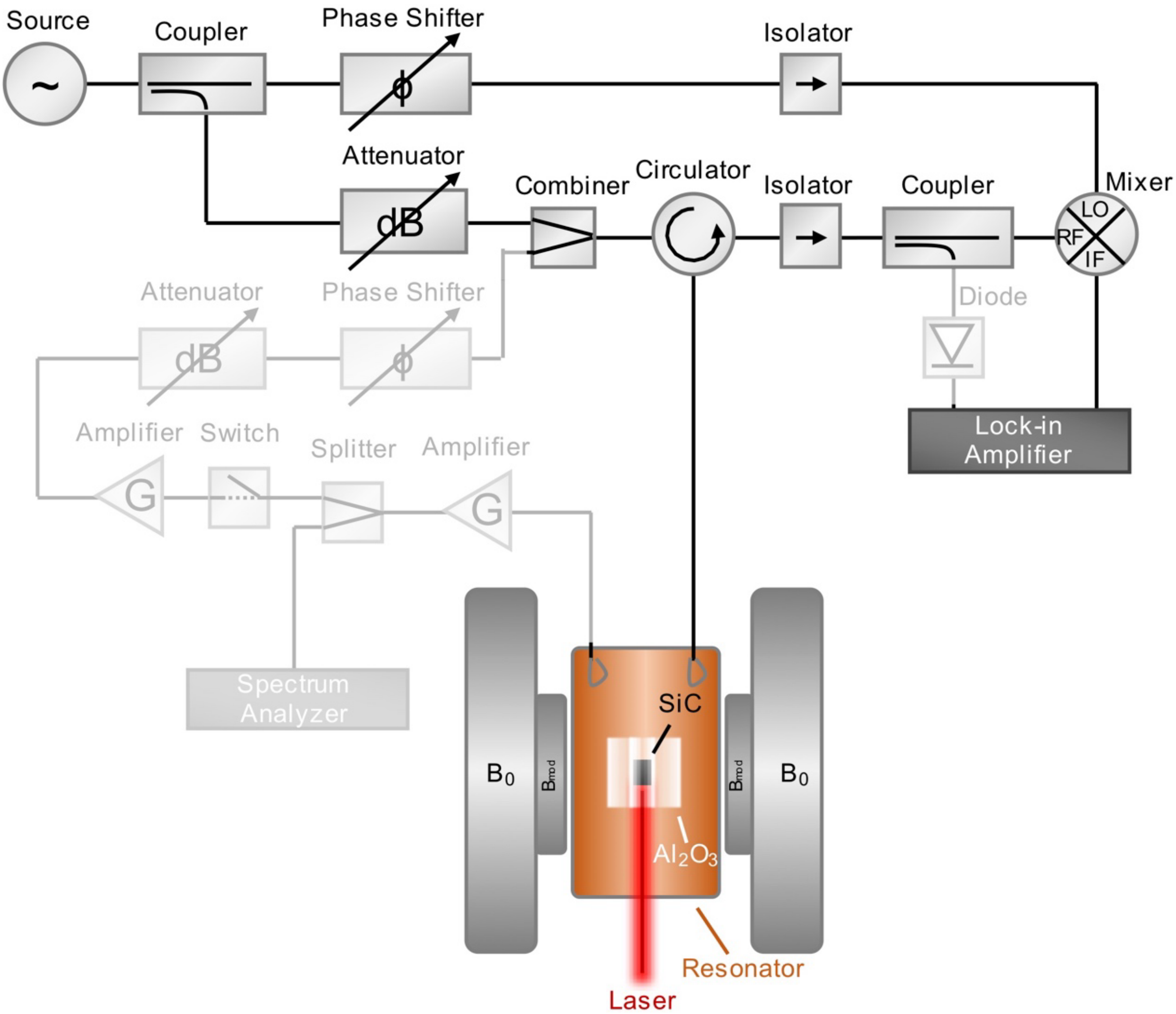

**Supplementary Figure 3. Setup for electron paramagnetic resonance.** The microwave frequency is fixed to the resonance frequency of the resonator. The main part of the microwaves drives phase-adjusted the local oscillator of a mixer. A small part is attenuated to a sufficiently low power and interacts with the sample. The remaining microwaves are detected via the mixer and the lock-in amplifier.

With the previous measurement we extracted the correct frequency regime ($f \approx f_R$). Besides the frequency, a maser/amplifier/refrigerator requires also the correct magnetic field range ($B \approx B_{\pm}$). A perfect measurement for probing the resonance condition of the spin system is electron paramagnetic resonance (EPR). Again, the microwave reflection is measured, but the swept quantity is the magnetic field using a pair of Helmholtz coils (see Supplementary Fig. 3). The microwave output is fixed to the resonance frequency $f_R$ in a continuous wave mode. To enhance the signal-to-noise ratio we are using the same lock-in amplifier (second input), and the magnetic field is sinusoidally modulated with a pair of small modulation coils. The interacting microwaves are attenuated (0 – 60 dB) to avoid saturation of the spin system.

The main part of the microwaves is guided through the first coupler to drive the local oscillator of the mixer. A phase shifter is implemented to phase-adjust the microwaves, thus providing the correct phase of the down-converted intermediate frequency. The remaining down-converted frequency is identical to the modulation frequency and recorded by the lock-in amplifier. Besides the information of the resonant magnetic field ($B \approx B_{\pm}$), EPR reveals the spin polarization of the system (see [1,2] for details). Thus, the pump threshold for the maser can be estimated with EPR.

### Maser

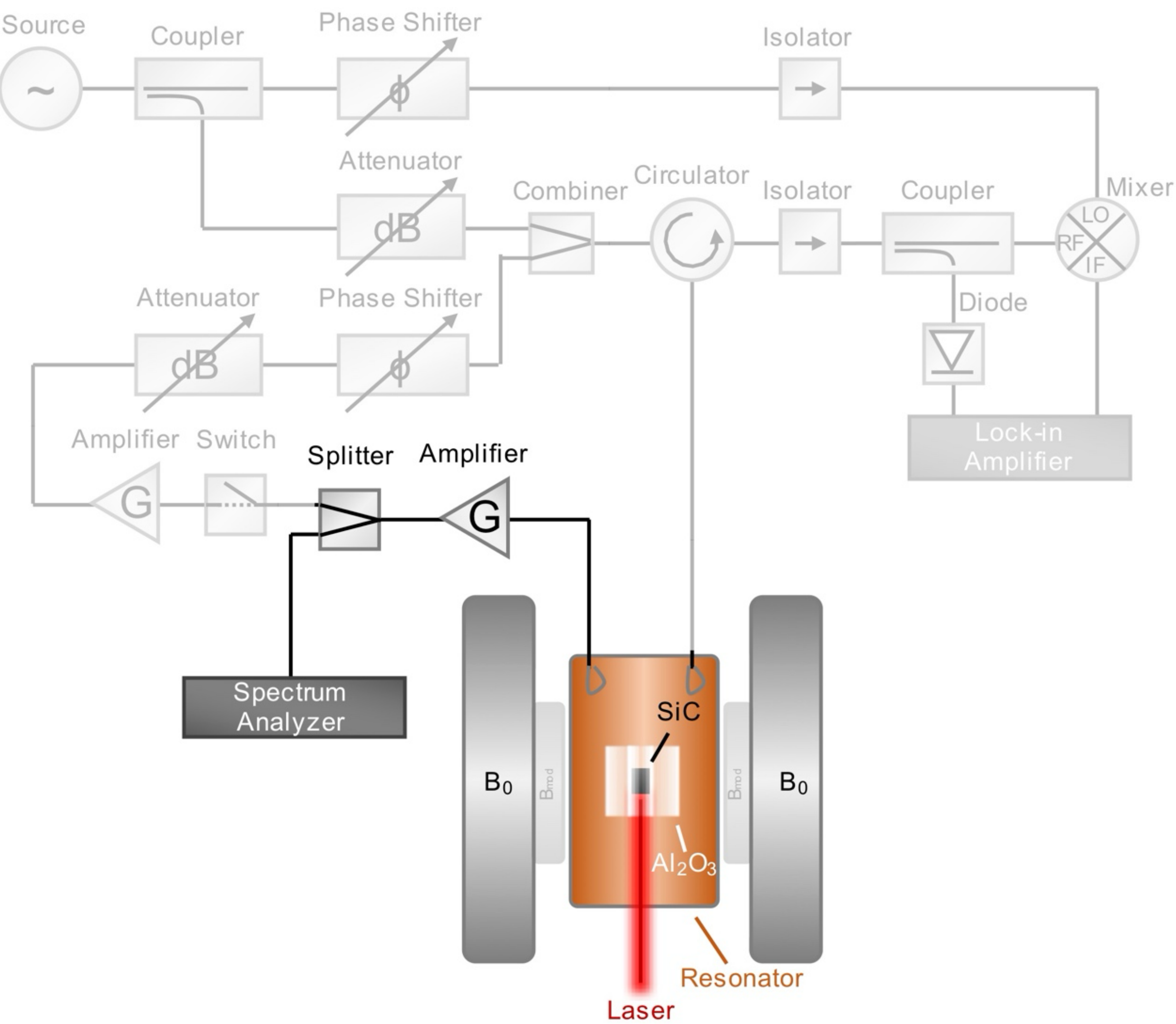

**Supplementary Figure 4. Setup for maser measurements.** The sample is excited via a laser within an external magnetic field. The emitted microwaves are amplified (to compensate losses) and guided to a spectrum analyzer.

For the maser output measurement, the magnetic field range is arranged according to the previous EPR experiment, and the frequency range of the recording spectrum analyzer is adjusted to the resonance frequency of the resonator. To compensate several losses (e.g. inserted splitter for the next measurement type), a LNA is inserted.

**Q-boosted Maser**

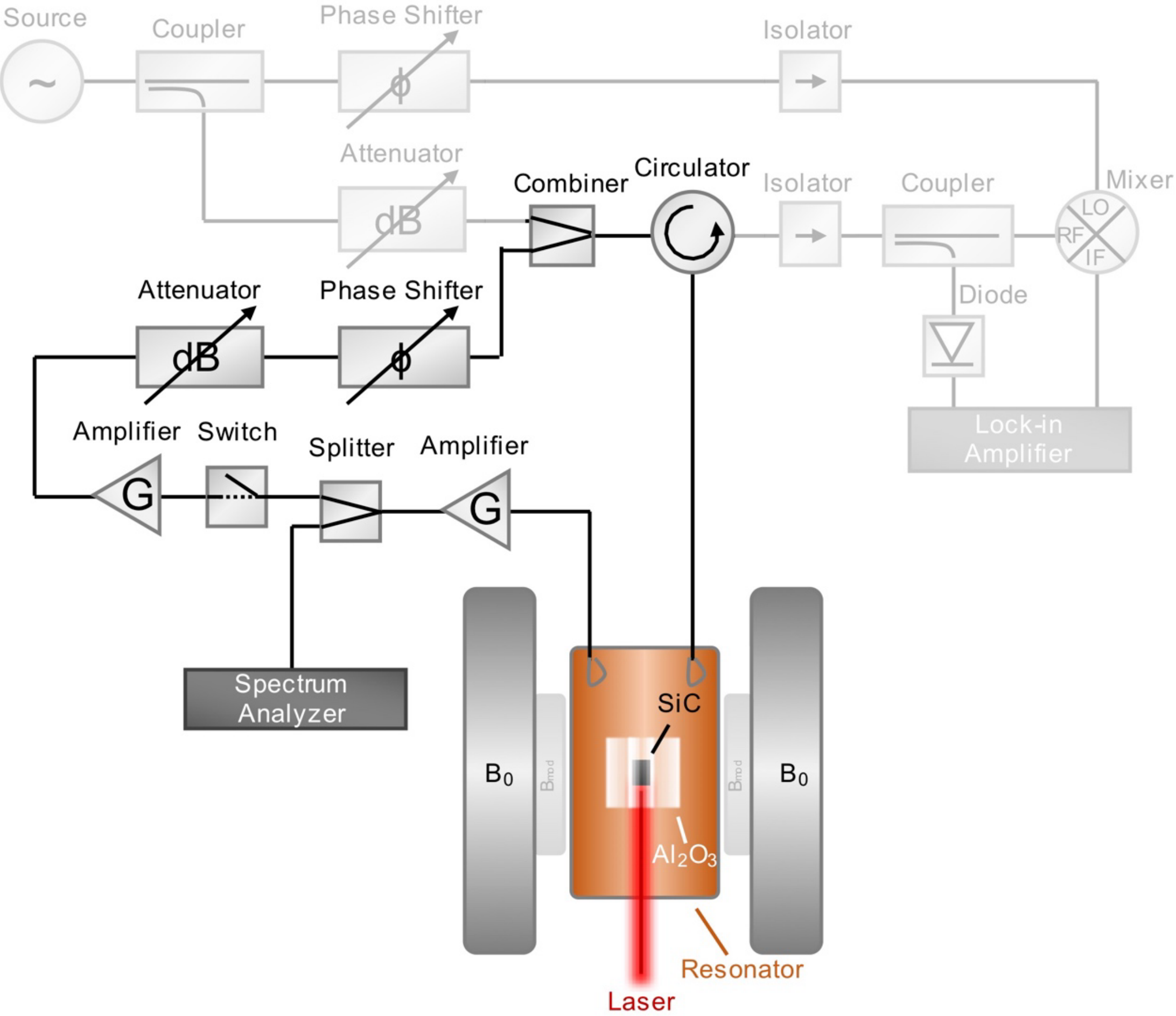

**Supplementary Figure 5. Setup for Q-boosted maser measurements.** By closing the microwave switch, the feed-back loop can be activated. After amplification (+attenuation) and phase adjustment the microwaves are inserted back into the resonator. Only a part of the signal is guided to the spectrum analyzer.

The artificial enhancement of the Q-factor is realized by a feed-back loop. The Q-boosting is activated by closing the microwave switch. 50% of the output signal after the LNA are amplified by a second amplifier. To avoid an overdrive the gain is controlled by attenuating (0 – 60 dB) the previously amplified signal. After phase adjustment via a phase shifter, the microwaves are inserted back into the resonator.

### Amplifier/Refrigerator

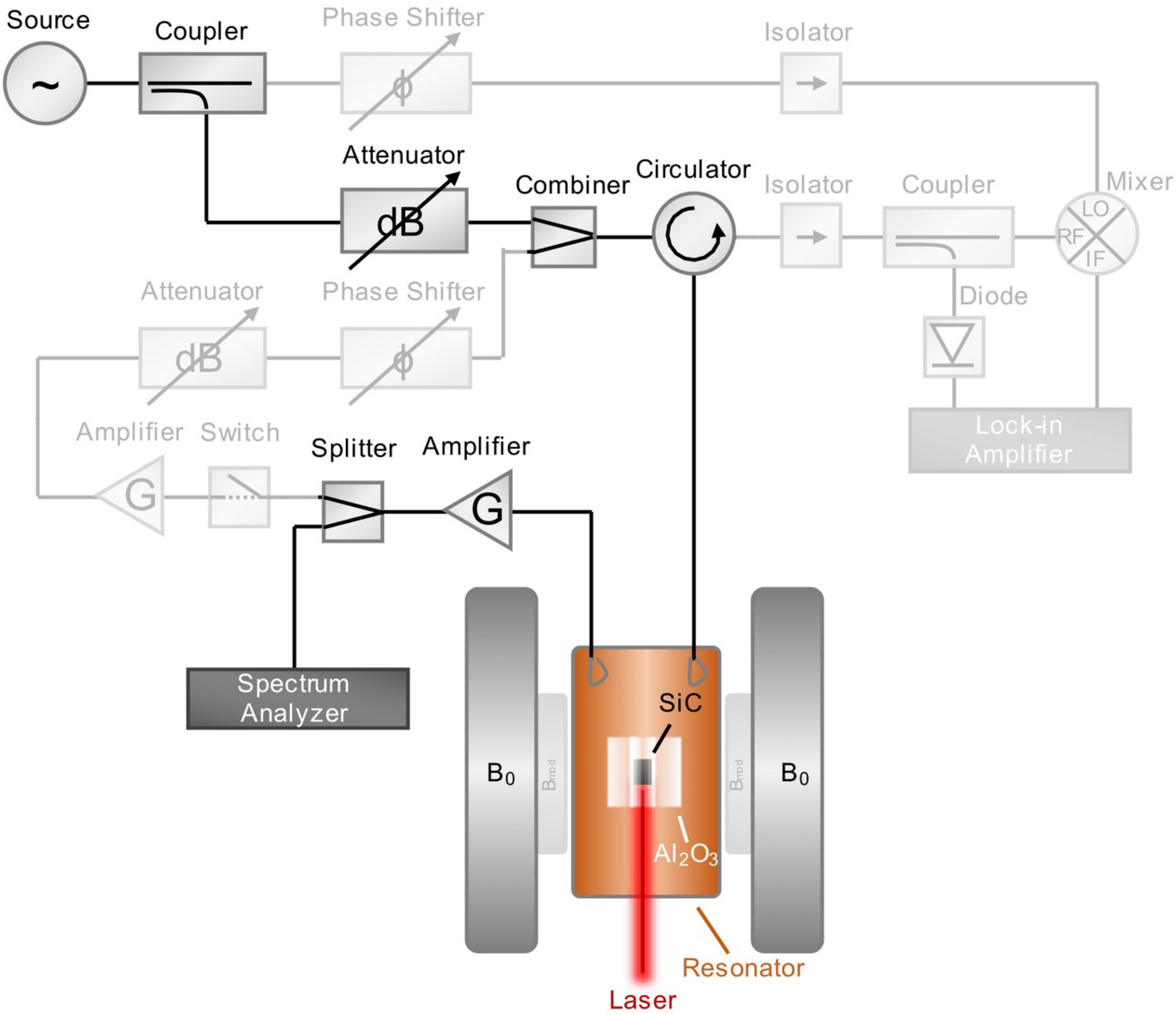

**Supplementary Figure 6. Setup for amplifier/refrigerator measurements.** An arbitrary microwave signal is inserted into the resonator by one antenna. A second antenna detects the output signal. The amplifier (refrigerator) can be switched on for $B = B_+$ ($B = B_-$) and off for $B \neq B_+$ ($B \neq B_-$), respectively.

The amplifier and refrigerator are analyzed by inserting a test signal into the system. The frequency is adjusted to the resonance frequency and the magnetic field within the range of the resonant condition. The amplifier (refrigerator) can be switched off by an off-resonance magnetic field $B \neq B_+$ ($B \neq B_-$). As soon as the magnetic field is in the range of the resonance condition $B = B_+$ ($B = B_-$), the spin system interacts with the microwaves, and the amplification (absorption) is recorded with the spectrum analyzer.

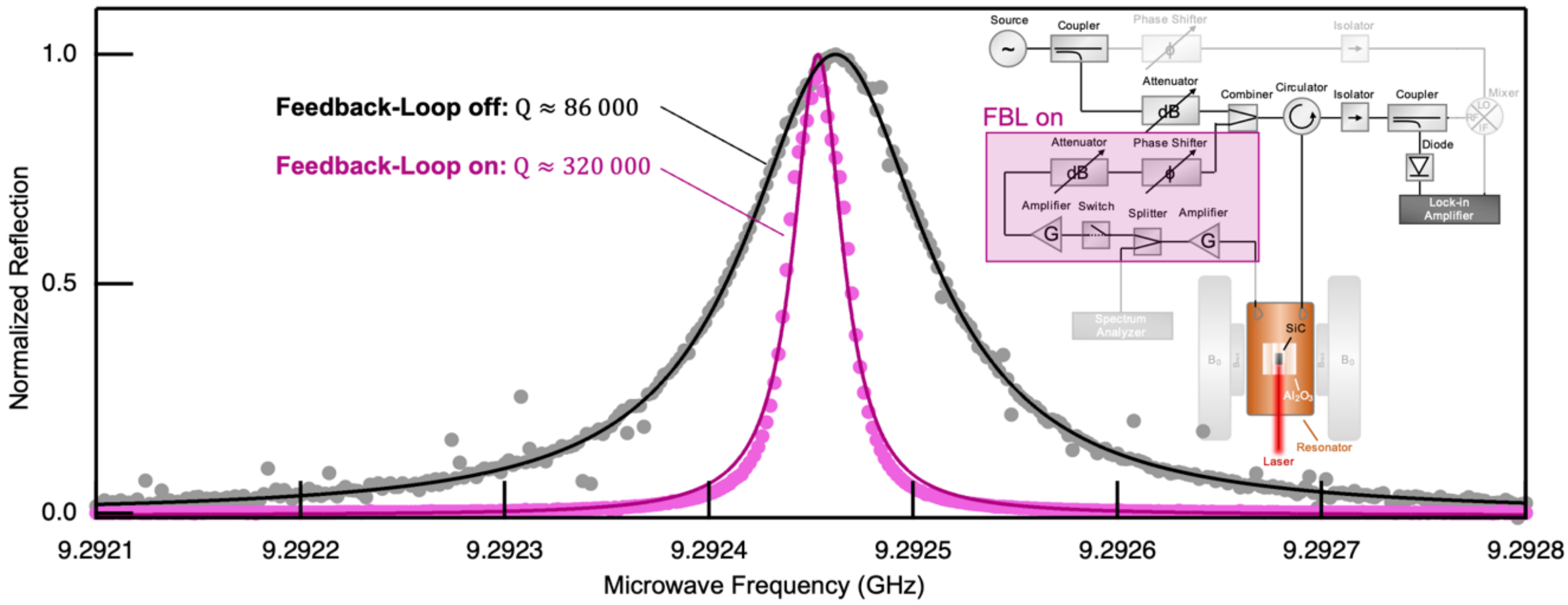

**Supplementary Figure 7. Effect of the activated feedback loop on the Q-factor.** A portion of the extracted microwave signal is amplified and fed back into the system to interfere constructively with the circulating microwaves. A phase shifter within the loop ensures constructive interference, while a variable attenuator positioned after the amplifier controls the overall gain of the feedback loop.

## Resonator Design

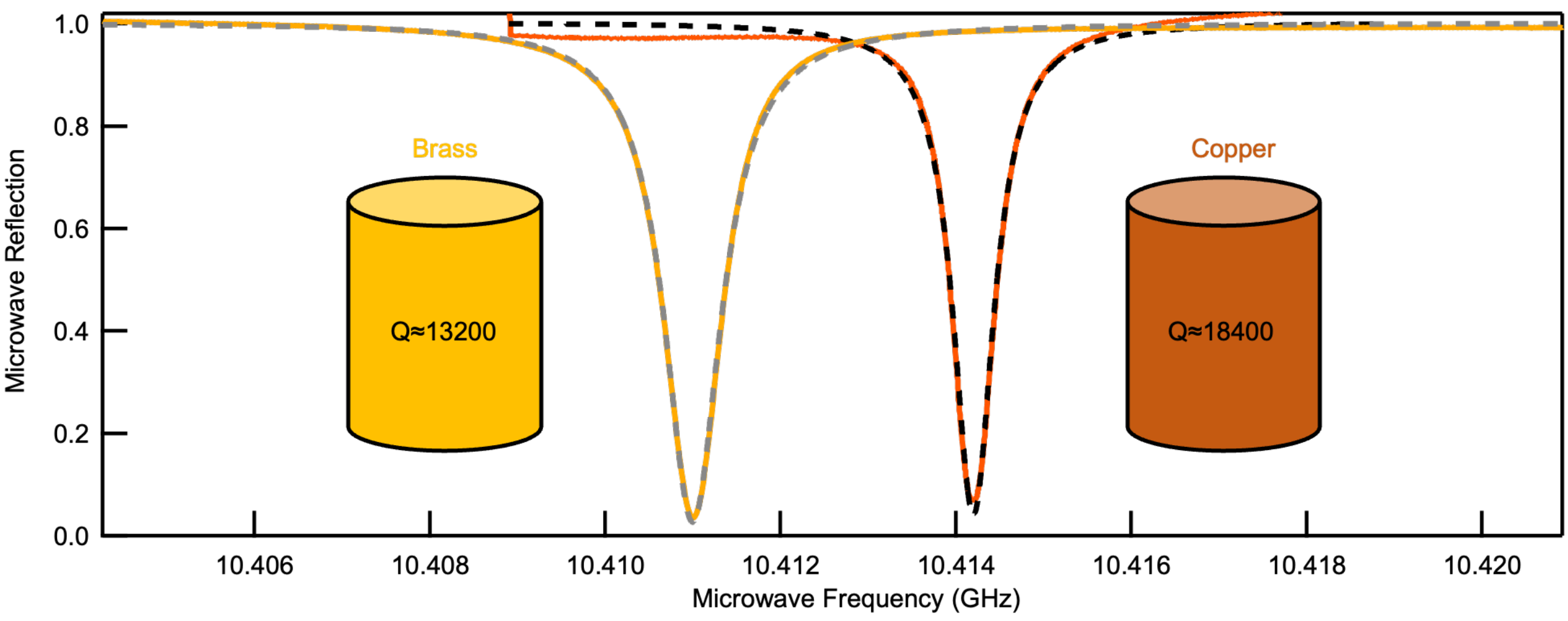

**Supplementary Figure 8. Influence of the resonator material on the resonator Q-factor.** Due to the higher conductivity, the ohmic losses are reduced in a copper resonator in comparison to a brass resonator with identical dimensions. The Q-factor values are shown here for an empty cavity (with sapphire, but without silicon carbide).

Different parameters were varied iteratively to find the most suitable resonator for the maser/amplifier/refrigerator. We started with an empty brass and copper resonator with dielectric sapphire core and compared the influence of conductance on the Q-factor (see Supplementary Fig. 8). As expected, the copper resonator provides a higher value. Therefore, we continued with height-dependent measurements (see Supplementary Fig. 9 a and b). Here, we already included the gain material to find the best constellation for a SiC-based device. We further changed the diameter of the resonator with the highest Q-factor from the height-dependent measurements to get the highest value of Q = 17600 (see Supplementary Fig. 9). To increase the Q-factor even more we can either under couple the system or cool down the resonator to reduce ohmic losses (see Supplementary Fig. 10). The highest value of $Q_{max}$ = 85000 was reached for a temperature of $T = 87$ K. This was approximately the lowest possible temperature of the liquid nitrogen cryostat, since the whole ensemble has to be cooled resulting in thermal losses. However, as soon as the laser is activated the temperature stabilizes around $T = 110$ K.

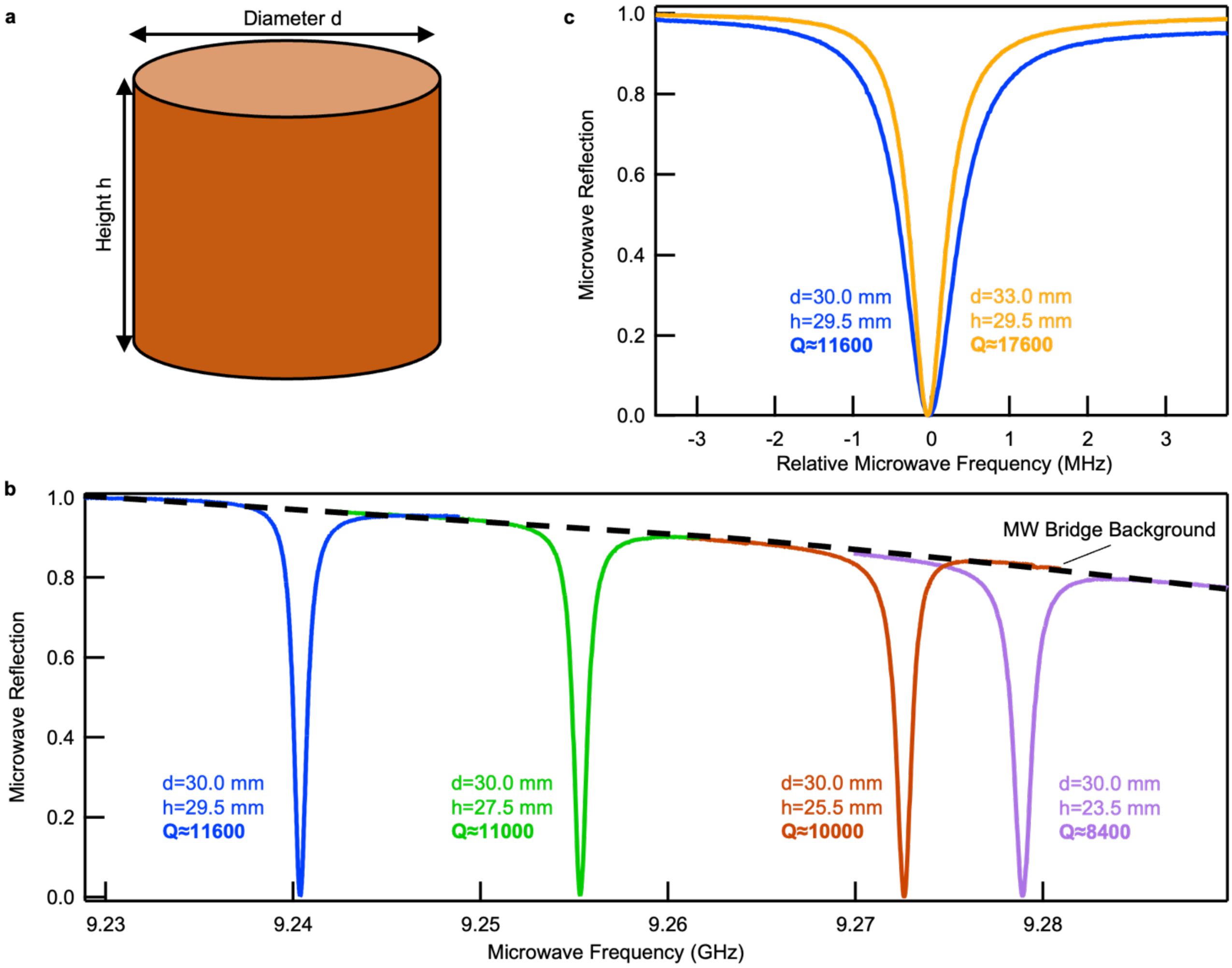

**Supplementary Figure 9. Influence of the resonator dimensions on the Q-factor. a** The height $h$ is varied as well as the diameter $d$ of the copper resonator filled with silicon carbide in order to find the highest Q-factor. **b** First, the height is varied. With increasing height, the resonance frequency shifts towards lower frequencies (absorption due to the microwave bridge background is indicated with a dashed line). **c** We chose the resonator with the highest Q-factor and enhanced the value by changing the diameter (relative frequency used due to large resonance frequency shift of 78 MHz).

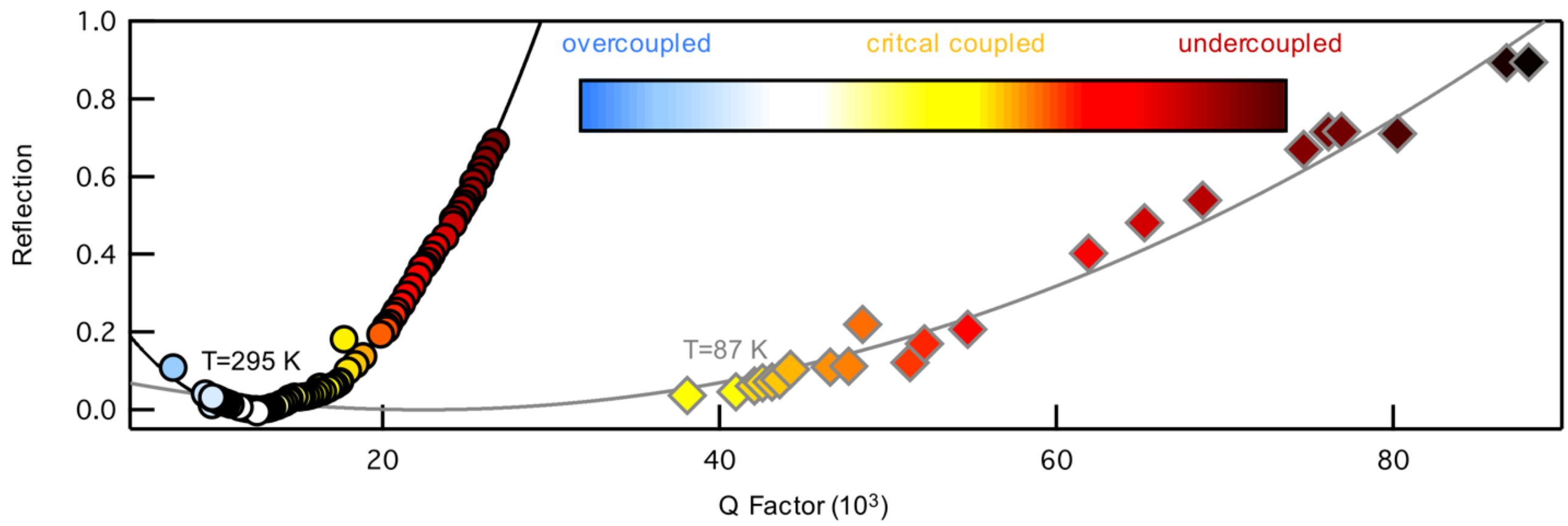

**Supplementary Figure 10. Quality factor variation of the resonator without feedback loop.** The Q-factor can be varied slightly by moving the coupling antenna in and out of the resonator. Depending on an over coupled (blue), critically coupled (yellow) and under coupled (red) state, the resonator reaches values up to 30000 at room temperature (black circles) and 85000 at cryogenic temperatures (grey diamonds). To study resonators with even higher values an artificial enhancement (feedback-loop) is required.

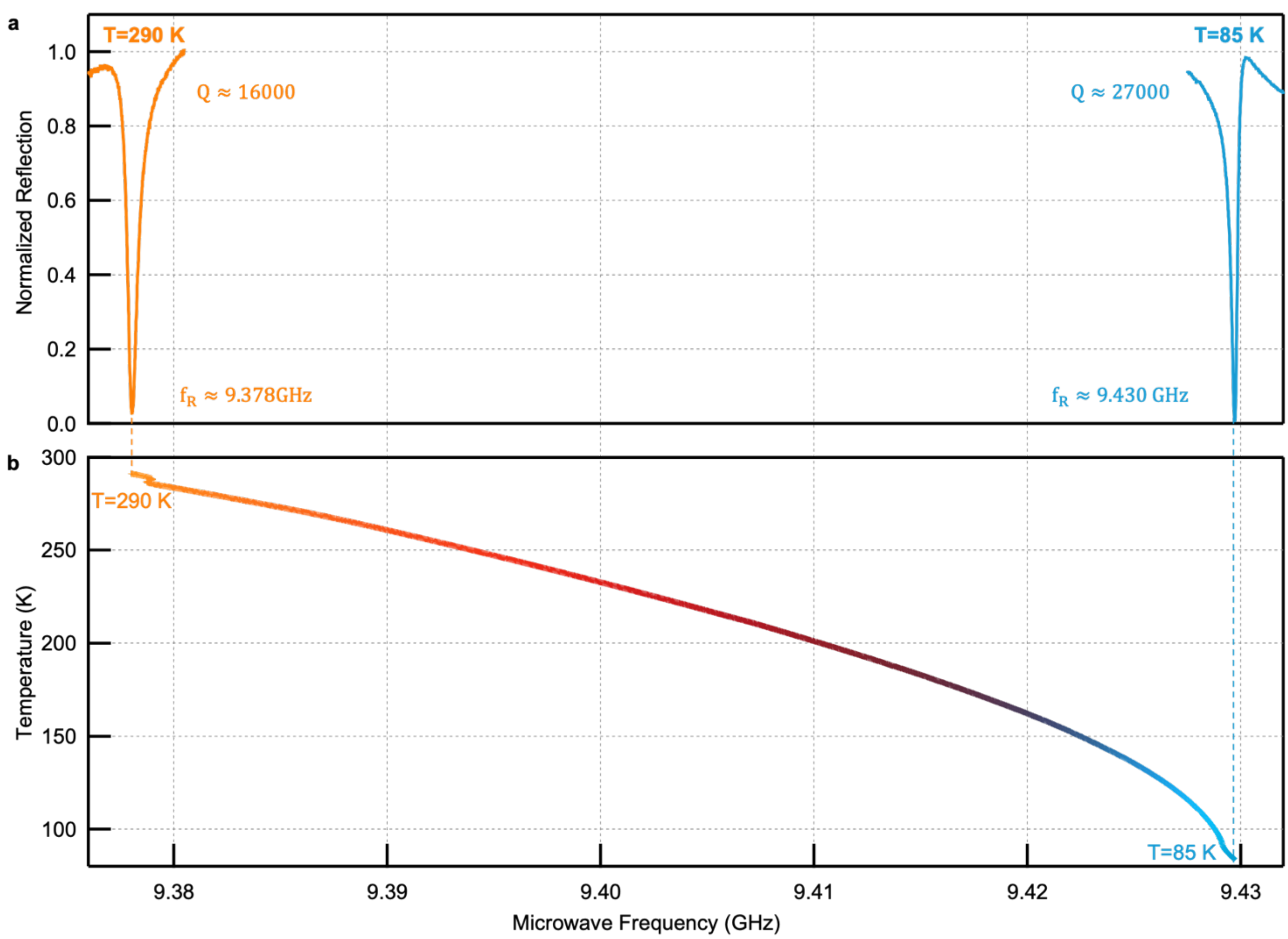

**Supplementary Figure 11. Temperature dependence of the resonator.** The resonance frequency is shifting by approximately 50 MHz within a temperature range from 85 – 290 K due to thermal expansion. The Q-factor is measured at 290 K and 85 K. Between the two resonances, an automatic frequency control locks the resonance frequency of the resonator and provides intermediate frequencies for the whole temperature range.

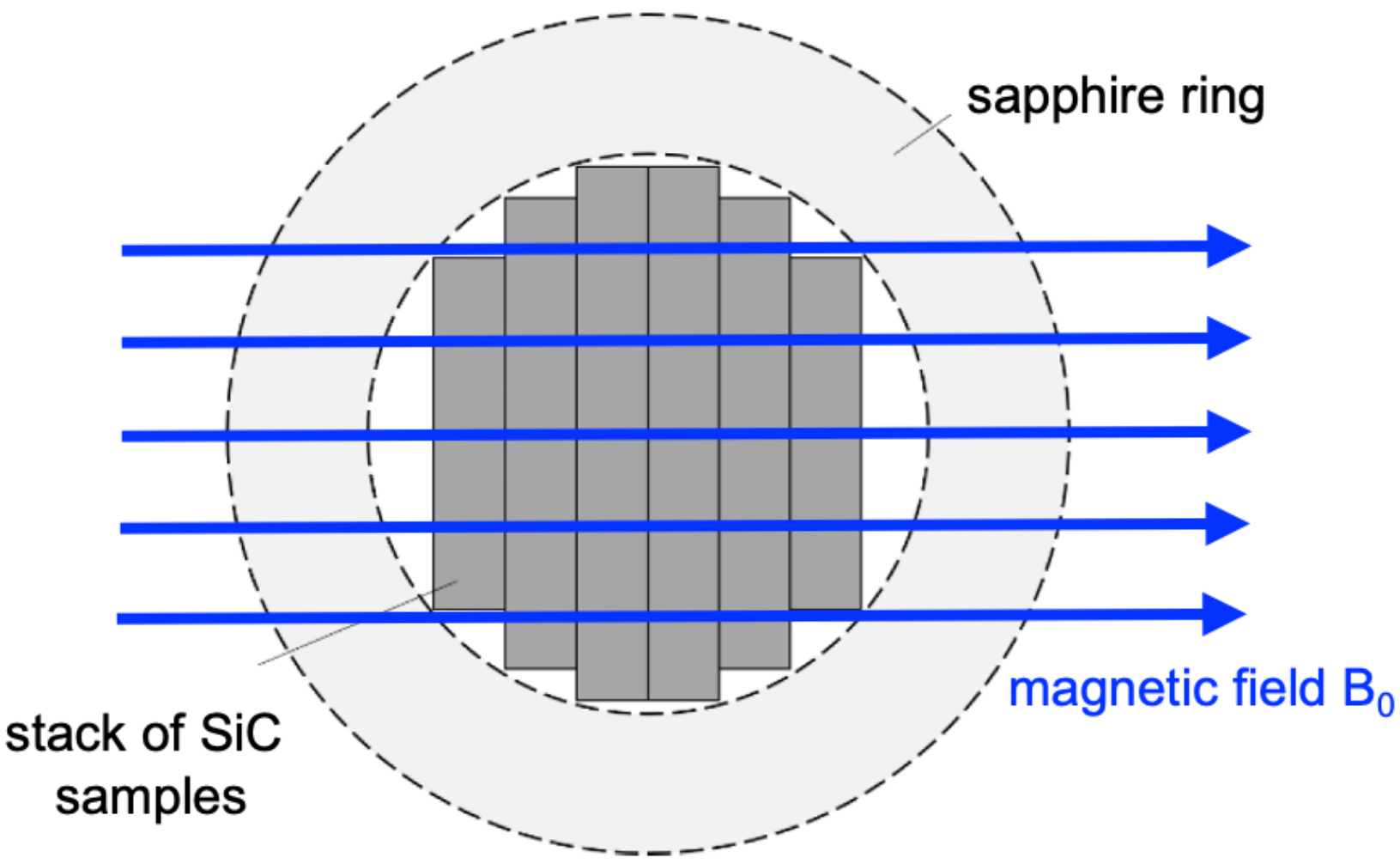

**Supplementary Figure 12. Sample geometry within the resonator.** In order to achieve a high filling factor we use a stack of rectangular SiC samples. Notably, the size varies to fit into the cylindrical sapphire ring. We chose this rectangular design instead of having round SiC discs to ensure a parallel magnetic field with respect to the crystal axis. The magnetic field would be perpendicular for SiC discs, leading to a 50% population inversion reduction.[3]

### Maser/Amplifier simulation and experimental data

Simulations according to [4] were performed with the following parameter set:

| Parameter | Value | Source | Diamond Maser Values [7] |
|---|---|---|---|
| $\gamma_{eg}$ | 26 $s^{-1}$ | $T_1^{-1}(T = 110\text{ K})$ | 208 $s^{-1}$ |
| $N$ | $2.19 \cdot 10^{13}$ | see Figure S11 | $4 \cdot 10^{13}$ |
| $\omega_c$ | $2\pi \cdot 9.3$ GHz | Q-factor measurement | $2\pi \cdot 9.2$ GHz |
| $T_2^*$ | 250 ns | based on EPR | 500 ns |
| $g$ | $2\pi \cdot 0.05$ Hz | see Figure S12 | $2\pi \cdot 0.11$ Hz |
| P | 1 – 10000 mW | laser power range | 400 mW |
| $Q$ | $10^4$-$10^6$ | FB-loop range | $3 \cdot 10^4$ |

$\gamma_{eg}$: The value of 26 $s^{-1}$ is extracted from [5] for a temperature of $T = 110$ K.

$N$: The number of participating spins is determined by dark EPR. The corresponding measurement is depicted in Figure S13. A piece of the SiC wafer (irradiated with 2MeV electrons with a fluence of $2 \cdot 10^{17}$ $cm^{-2}$) was investigated in a Magnettech ESR5000 at room temperature (Figure S13 a). We used the left transition (Figure S13 b) and compared the double integral (area of the bottom plot) with the value of a reference sample with a known spin number (BDPA in Supplementary Fig. 13c: $n = 3.6 \cdot 10^{17}$). Thus, we extracted a spin number of $2.7 \cdot 10^{12}$ for the SiC sample resulting in a spin density of $2.27 \cdot 10^{15}$ $cm^{-3}$. The total volume of the sample used for the maser/amplifier/refrigerator is 0.0344 $cm^3$ resulting in a total spin number of $7.8 \cdot 10^{13}$. The spin number for one transitions (see Figure 1b) is reduced to $\frac{1}{2} \cdot 56.1\% \cdot 7.8 \cdot 10^{13} = 2.19 \cdot 10^{13}$ taking the two transitions ($B_+$ and $B_-$) and the isotopic ratio of silicon (see [3] for detailed discussion) into account.

$\omega_c$: The value of 2π · 9.3 GHz is based on Q-factor measurements. Small variations can be observed between the measurements, however, the difference is in the MHz regime.

$T_2^*$: To access the coherence time in the setup used here, we analyzed the linewidth of the system with continuous-wave (CW) electron paramagnetic resonance (EPR). Notably, pulsed EPR measurements are not applicable here, as the Q-factor is too high for such experiments, leading to a long ring-down time of several microseconds. Therefore, we determined the $T_2^*$ time using the natural linewidth. We observed a linewidth of approximately 45 µT, which can be attributed to $T_2^*$ as there is no evidence of saturation due to overmodulation or power broadening. This corresponds to a frequency linewidth of 1.3 MHz, which in turn indicates a coherence time of approximately 250 ns. This value agrees with those reported by [6], where a $T_2^*$ time of 200 – 300 ns was observed in natural SiC at a similar irradiation dose.

$g$: Based on the resonator design, the coupling strength is expected to lie within the range $g/(2\pi) =$ 0.02 Hz – 0.11 Hz. [4] This value is influenced by numerous parameters and is not straightforward to determine directly. Therefore, we treated it as the sole free parameter in our model and varied it accordingly. As a reference point for the maser threshold, we used the maser measurement with the lowest output power (Figure 1c), since all other measurements yielded significantly higher maser outputs. By varying the coupling $g$, we observed corresponding shifts in the maser threshold along the Q-factor axis. Representative examples are shown in Supplementary Fig. 15. The best agreement with the experimental data was obtained for $g$ = 2π · 0.05 Hz, which is consistent with values reported by [4].

P: The value of the laser power is swept in the range of 1 mW and 10 W since this is within the experimental achievable range (*K808FANFA-15.00W* laser from *BWT*).

$Q$: The Q-factor axis was plotted for $10^4$ – $10^6$. Values of $10^4$ – $10^5$ are reached by an under coupling of the resonator (see Supplementary Fig. 9). Values exceeding $10^5$ are only possible via a feed-back loop which can be automatically activated and controlled by a microcontroller.

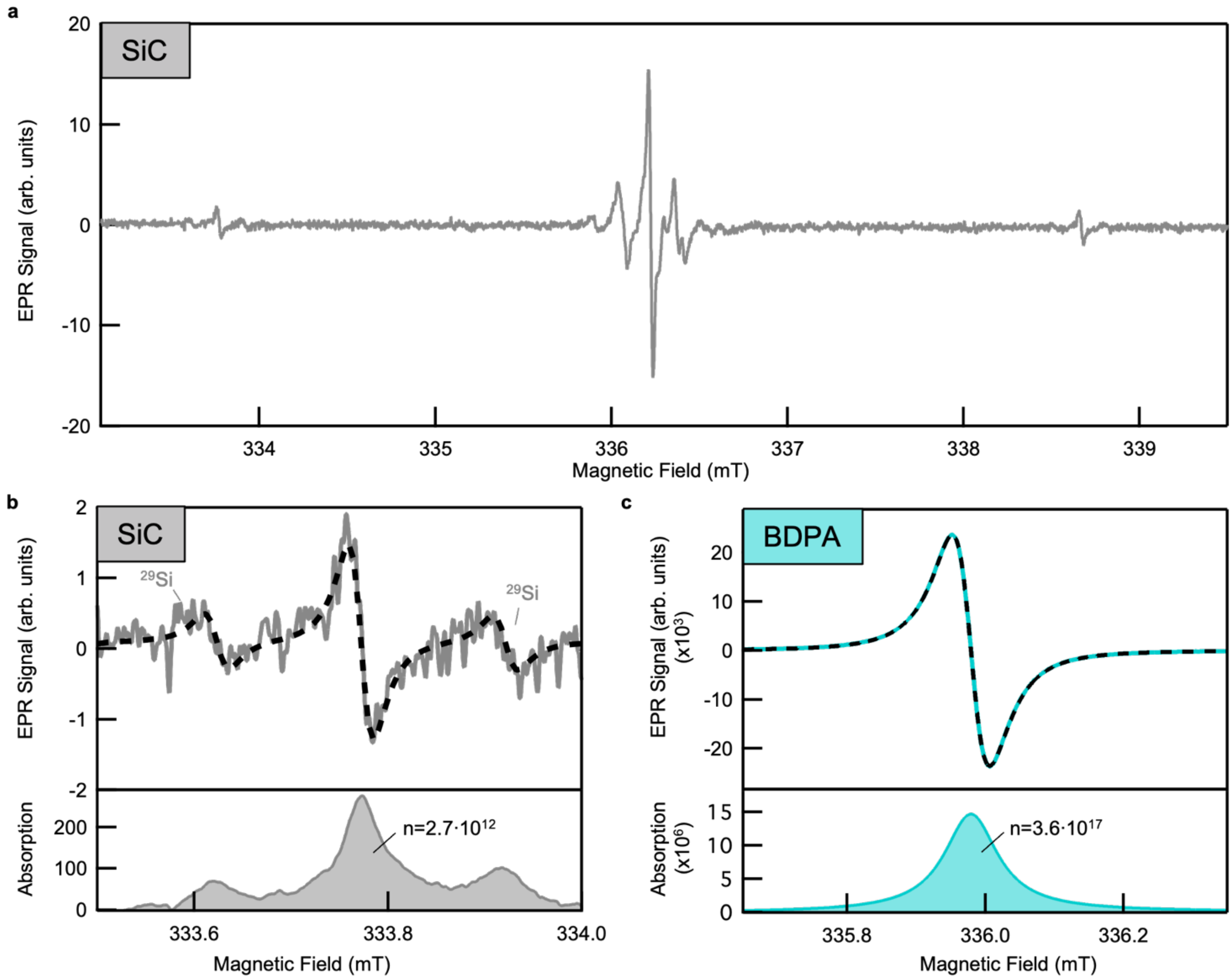

**Supplementary Figure 13. Quantitative EPR measurements determine the absolute spin number. a** Room temperature continuous wave EPR measurement without illumination of a SiC sample used in this paper. **b** Zoom into the left transition in order to get only the contribution of the V2 defect. The double integral is proportional to the absolute spin number $n$. **c** EPR spectrum of a reference sample (BDPA) with known spin number. All measurements were performed in a Magnettech ESR5000.

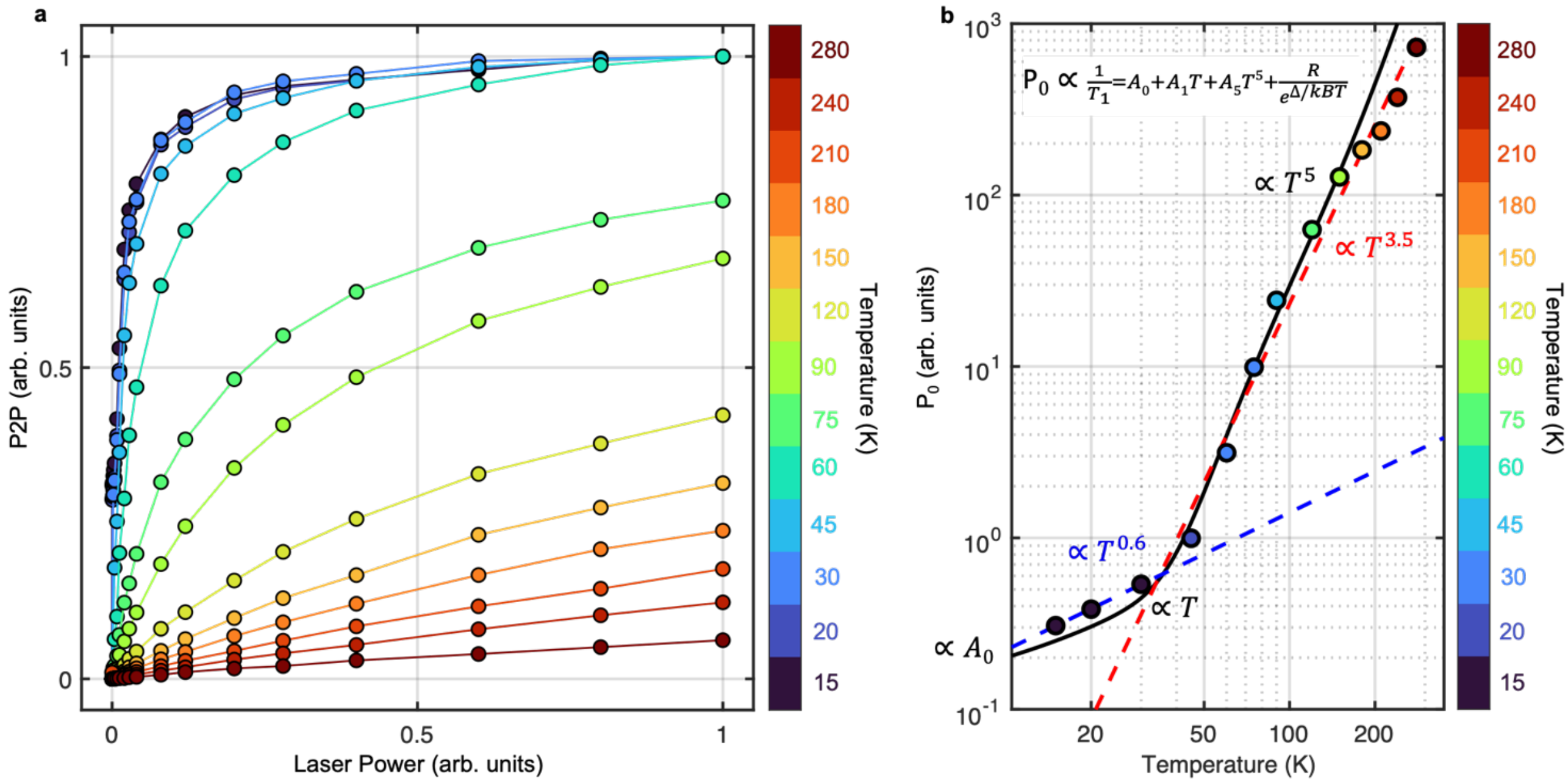

**Supplementary Figure 14. Temperature-dependent characteristic pump power measured via EPR** (see [1] for details). Measurements were performed using a home-modified Bruker E300. **a** Peak-to-peak amplitude of the EPR signal as a function of temperature and laser power. Fitting this behavior using parameters from [1] yields the characteristic pump power. **b** Extracted characteristic pump power as a function of temperature. The trend closely follows the known spin-lattice relaxation rate (shown in black) $\frac{1}{T_1} = A_0 + A_1 T + A_5 T^5 + \frac{R}{e^{\Delta/k_B T}}$ (see [5] for details). A quantitative analysis reveals minor deviations in the power-law dependence. Fits for the low- and high-temperature regimes are shown in blue and red, respectively, with their intersection marking the fit regime. At low temperatures, the regime approaches the $A_0$ plateau, which limits the fitting accuracy. At high temperatures, the dominant $\frac{1}{T_1} \propto T^5$ dependence is reduced due to the influence of continuous optical pumping on the relaxation time. Consequently, the temperature dependence is less pronounced.

## Further measurements and simulations

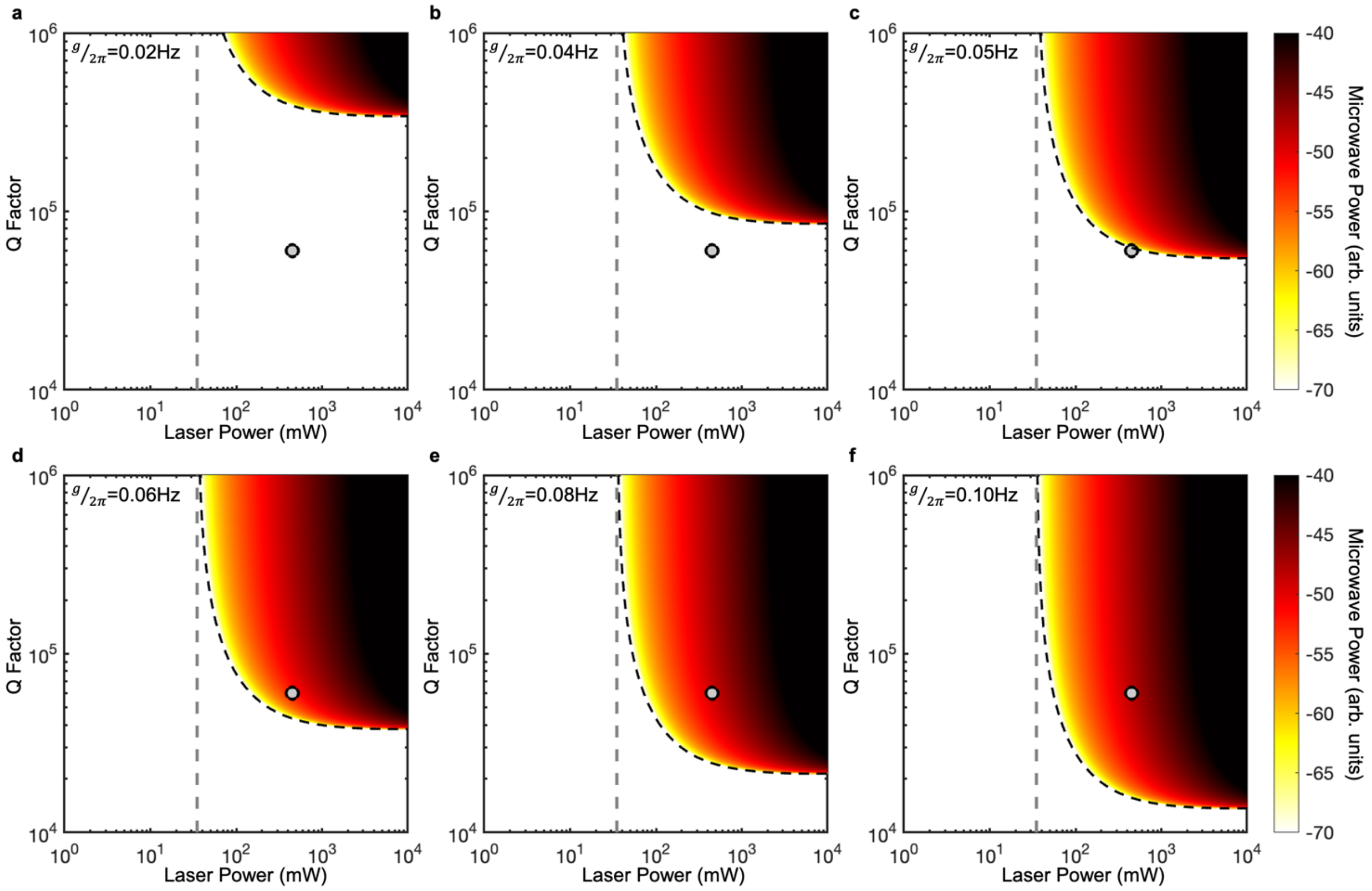

**Supplementary Figure 15. Maser simulations for different couplings $g$.** The grey circle illustrates the maser measurement of Figure 1c. Since the output power is very small, we expect the maser threshold to be in this parameter range. We achieve the best agreement with the model for $g/2\pi = 0.05$ Hz which is within the range of other reported values for similar resonators. [4]

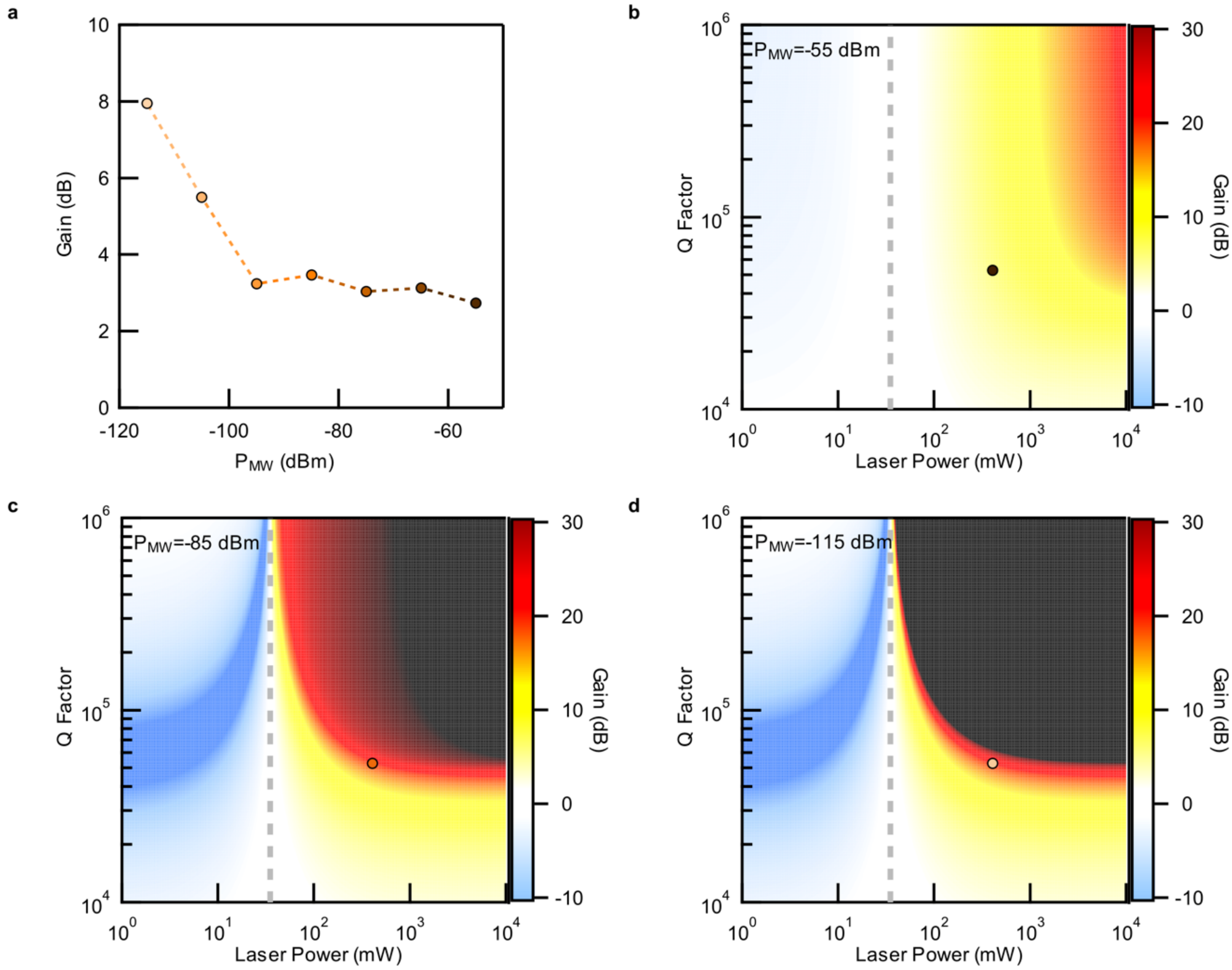

**Supplementary Figure 16. Power-dependent SiC-based amplifier. a** Measured gain for various input powers. **b-d** Simulations for different microwave input powers (-55 dBm, -85 dBm, -115 dBm). A higher gain is expected for smaller incoming signals, which fits qualitatively to the experimental observation.

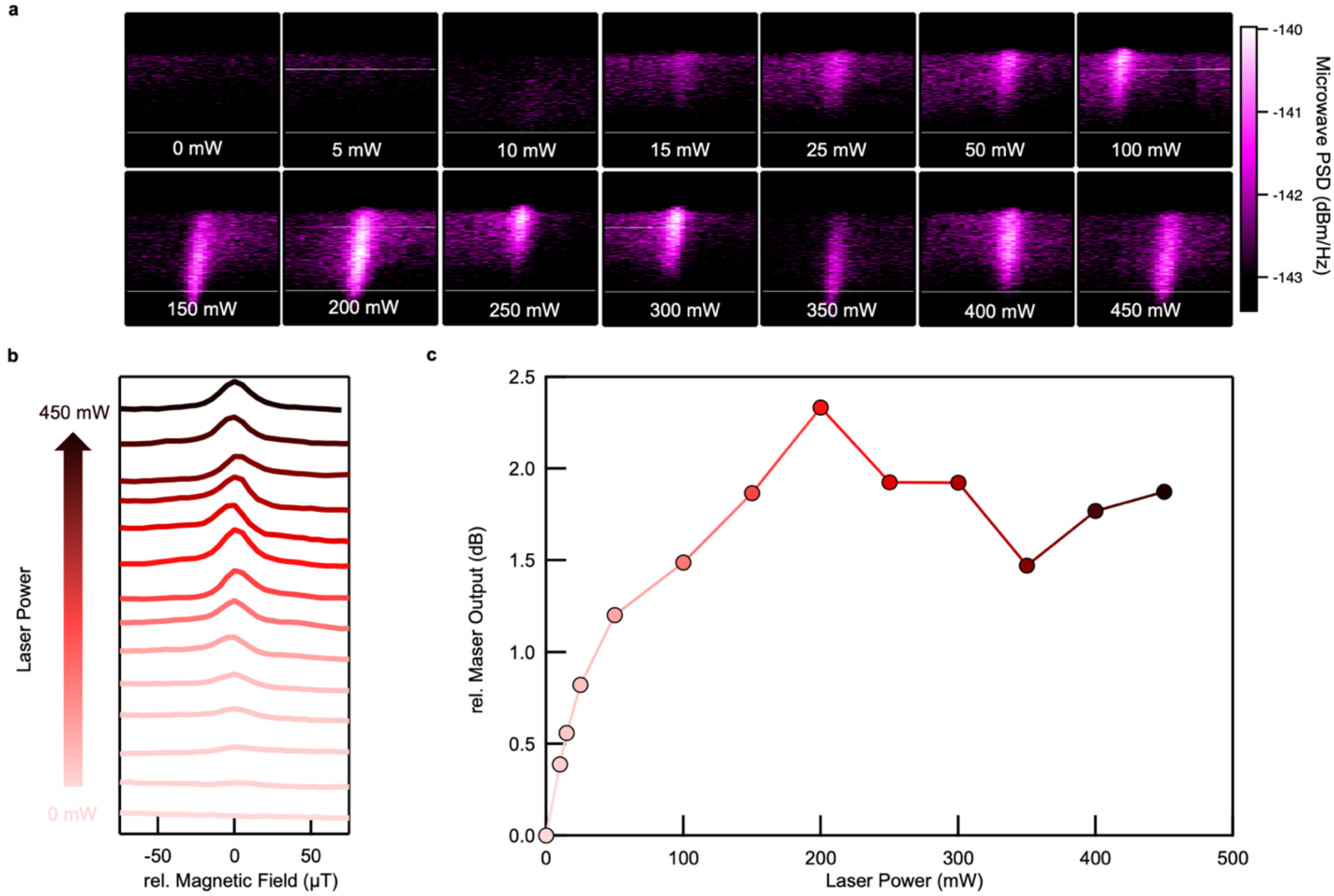

**Supplementary Figure 17. Laser power dependence of maser output and its saturation. a** Maser measurements without feedback loop at 110 K similar to Figure 1c in the main text. Masing occurs only at higher laser powers and saturates. **b** Qualitative analysis of the maser output by averaging across the different frequencies. **c** Quantitative analysis of the maser output for various laser powers with respect to the background noise. Saturation occurs for multiple hundreds of mW of laser power.

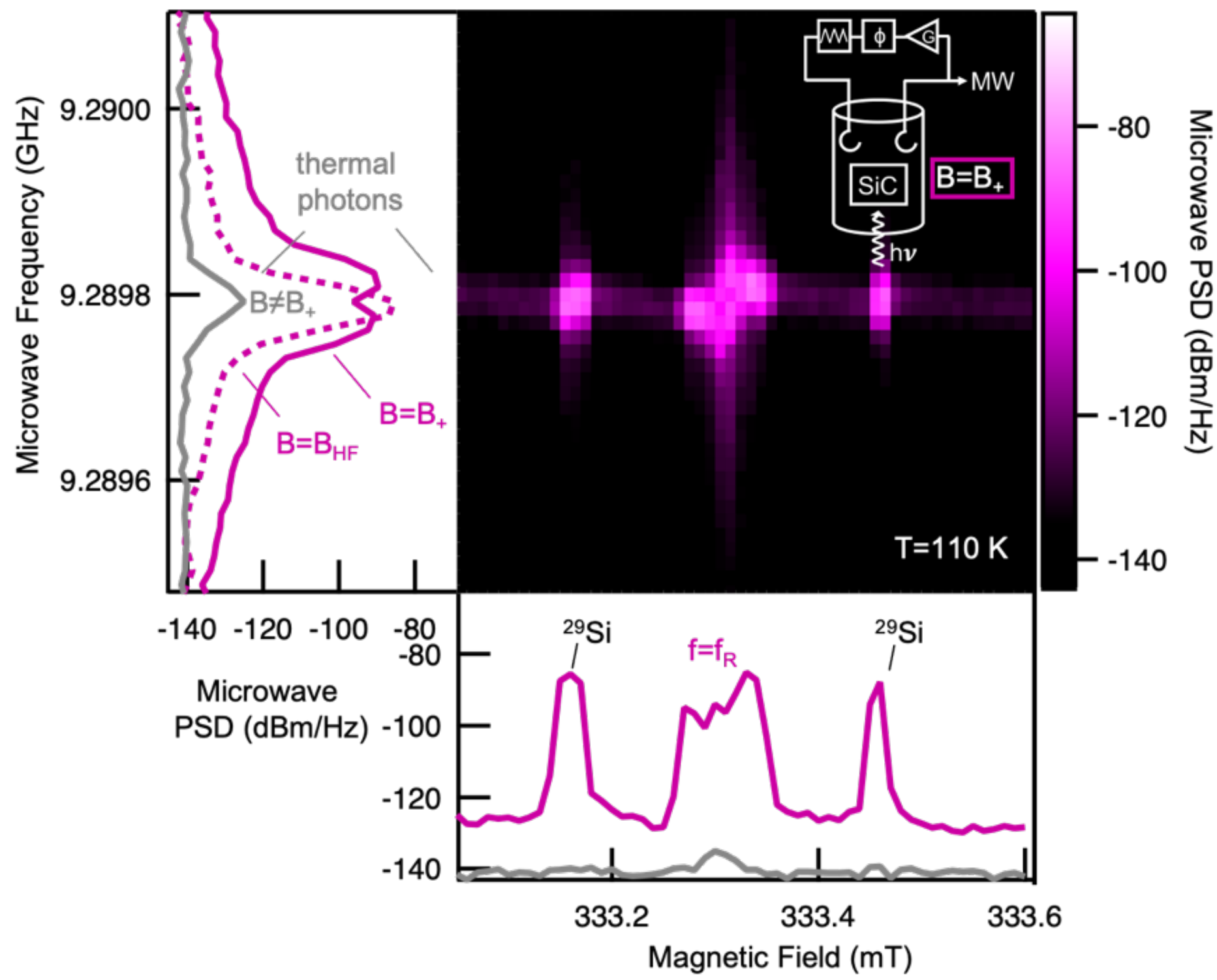

**Supplementary Figure 18. Maser saturation for high feedback-loop gain.** A strong indicator of saturation is the broadening of the main emission peak, accompanied by the saturation of the hyperfine peak intensities at the same intensity. Additionally, the presence of amplified thermal photons becomes evident, similar to the observation made at 315 K in the main text.

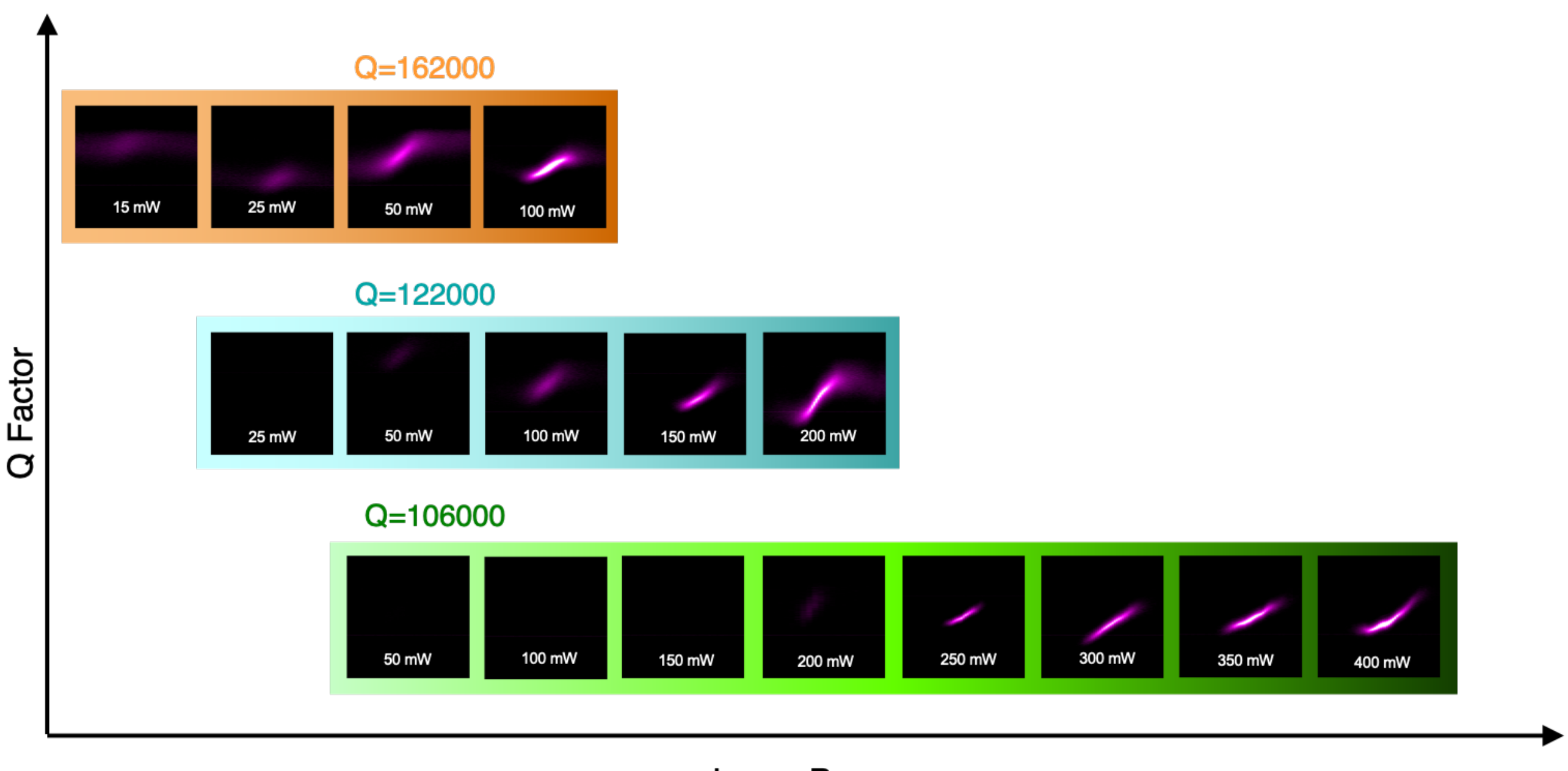

**Supplementary Figure 19. Overview of Q-boosted maser measurements for different laser power and Q-factors.** This figure summarizes the data sets which were used in the main text in Figure 2c. **a** Maser measurements for Q-factors of 162000 (orange), **b** 122000 (blue) and **c** 106000 (green). The color maps represent the microwave output power (pink) analog to Figure 2b in the main text. As expected from the simulations, masing occurs for higher Q-factors at lower pump powers.

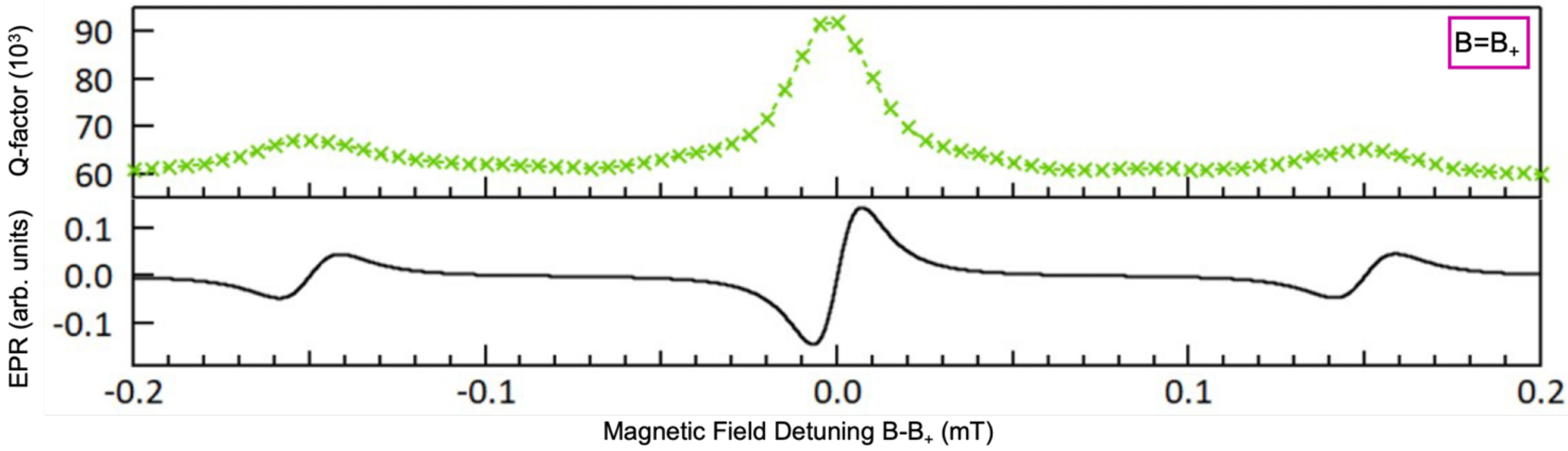

**Supplementary Figure 20. Impact of resonant transition on the Q-factor of the resonator.** All previously reported Q-factor values are off-resonant (in respect of magnetic field). Once the magnetic field is in resonance with the spin system, losses are reduced due to stimulated emission leading to an enhanced Q-factor of the system.

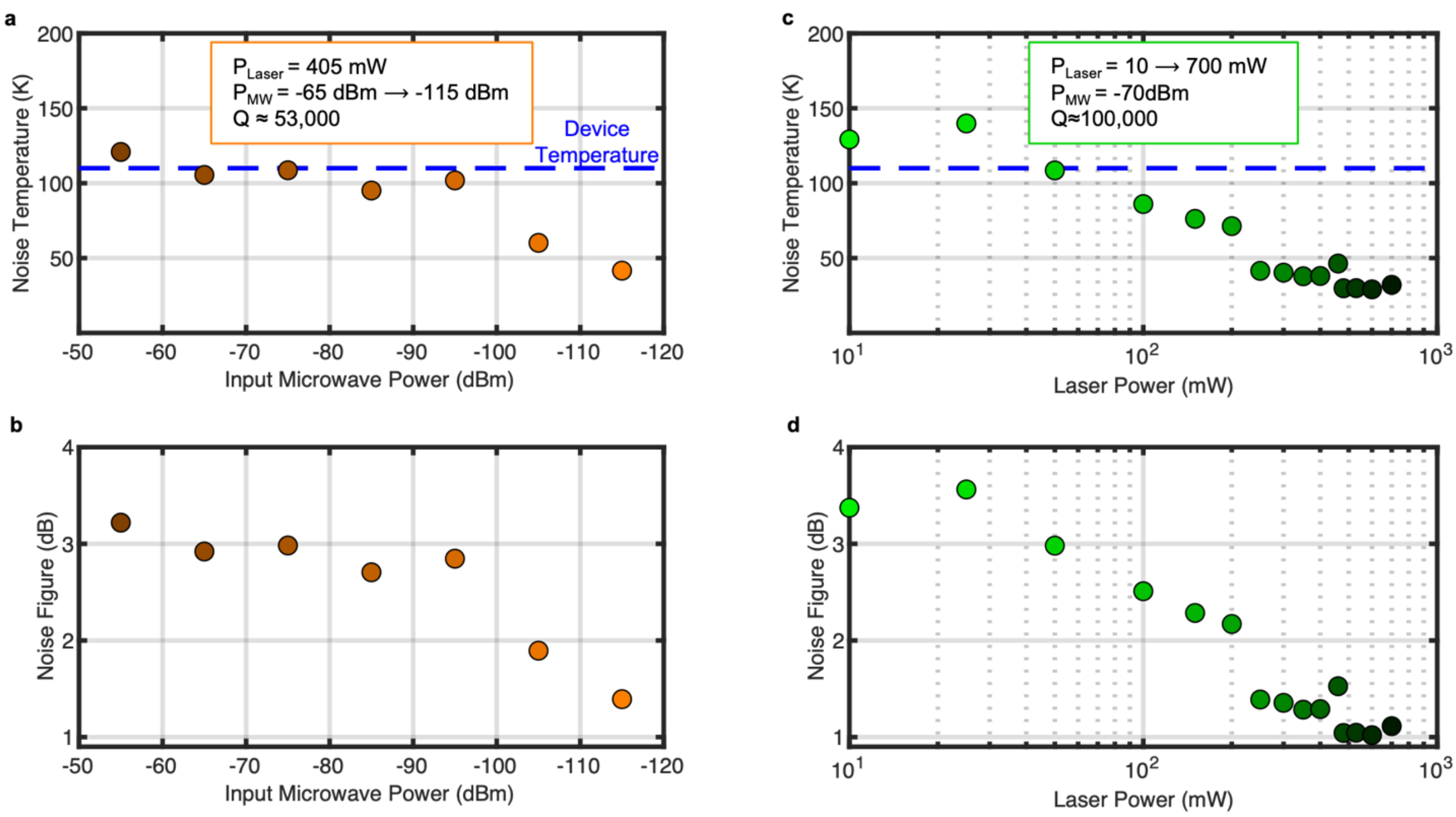

**Supplementary Figure 21. Noise performance of maser-based amplifier.** The noise temperature can be reduced below the physical device temperature by operating at low input powers and high laser pump powers. The noise figure ranges from 1 dB to 3.4 dB, following a consistent trend. Calculations are based on the model presented in [8].

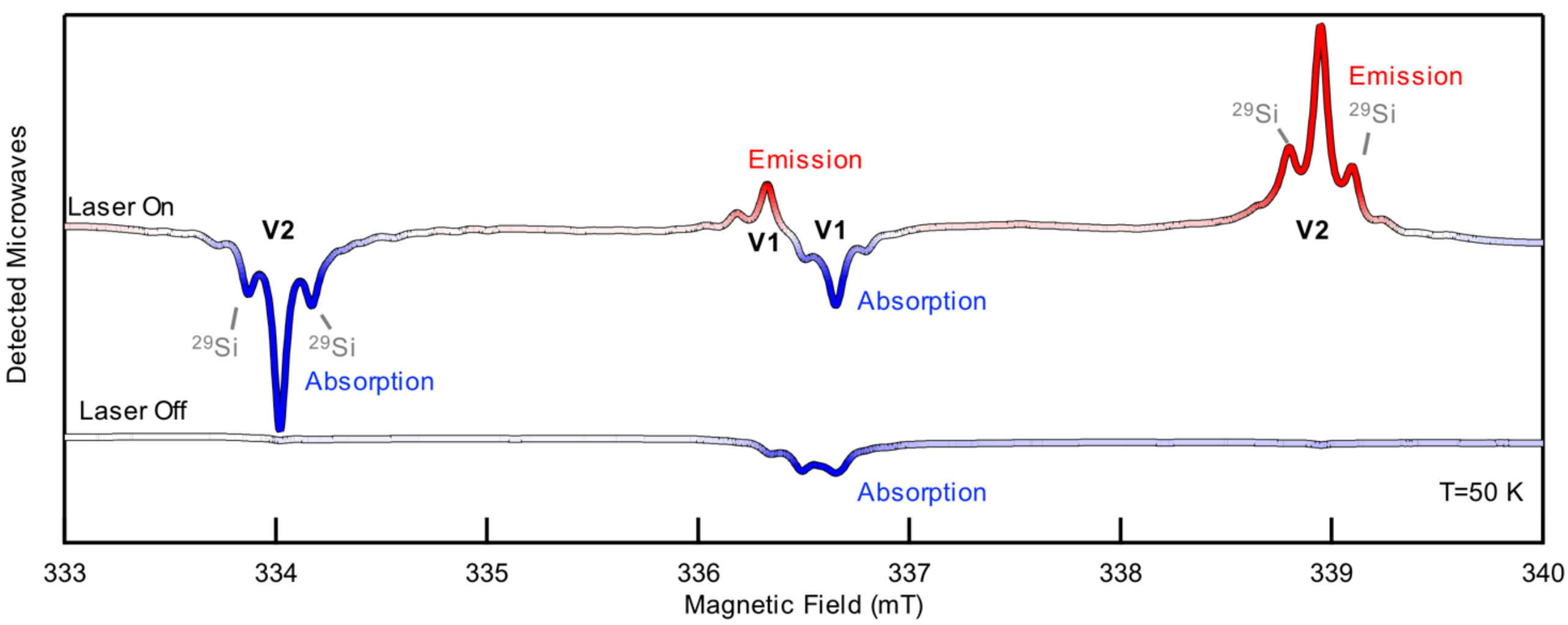

**Supplementary Figure 22. Comparison of the silicon vacancies V1 and V2.** The detected microwaves (integrated EPR signal) under illumination reveal the absorptive and emissive character of both defects at different magnetic fields. For proof-of-concept measurements, the V2 is favored since the effect of emission and absorption is larger and therefore easier to test for an amplifier/refrigerator. For future application of a magnetic switching of amplifier and microwave cooling devices (e.g. quantum computing) we recommend the V1 defect, since the field difference of the absorptive and emission transition is in the range of $\Delta B \approx 325$ µT due to the smaller ZFS.

### Estimation of the Magnetic Field Sensitivity

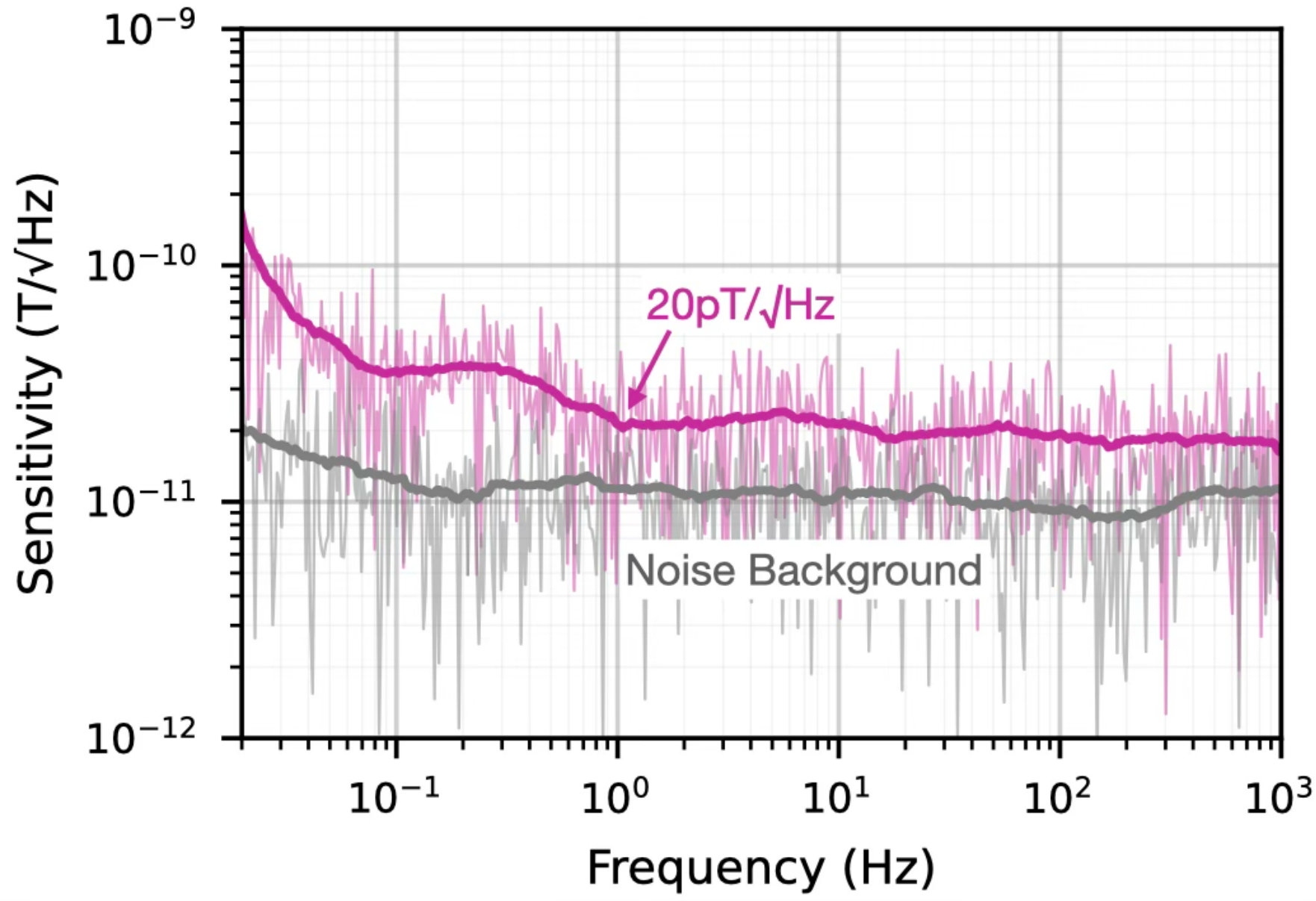

**Supplementary Figure 23. Magnetic-field sensitivity of the maser-based magnetometer.** The pink curve shows the laser-limited sensitivity estimate derived from the relative intensity noise of the excitation laser (808 nm, K808FANFA-15.00W, BWT), measured in current-sensing mode with a photodetector connected to a Tektronix DPO 3034 oscilloscope. The maser is operated well above threshold, such that small fluctuations around the operating point allow a first-order linear conversion of optical power noise into equivalent signal noise. Using the measured relative laser noise (in 1/√Hz) and the maser linewidth, we infer an estimated magnetic-field sensitivity of 20 pT/√Hz at 1 Hz. Slow temperature-induced drifts of the cavity occur on hour timescales and lie below the measurement bandwidth, while environmental magnetic and microwave-electronics noise are not included, as the figure represents a laser-limited estimate. The grey trace shows the oscilloscope background noise with the laser switched off.

In general, the magnetic field sensitivity of a magnetometer can be written as

$$\frac{\partial B}{\sqrt{\Delta f}} = \frac{\partial S}{\frac{\Delta S}{\Delta B}} \cdot \frac{1}{\sqrt{\Delta f}} = \frac{\Delta B}{\frac{\Delta S}{\partial S}} \cdot \frac{1}{\sqrt{\Delta f}}$$

where $S$ is the detected signal. For an ODMR magnetometer this would be the photoluminescence intensity, while in our case it is the emitted microwave power. For small variations around the resonance condition this expression can be approximated as

$$\frac{\partial B}{\sqrt{\Delta f}} \approx \frac{\Delta B}{\mathrm{SNR}} \cdot \frac{1}{\sqrt{\Delta f}}$$

Here, $\Delta B$ denotes the FWHM of the resonance. This follows directly from the linewidth definition: moving away from resonance by the HWHM reduces the signal amplitude to half of its maximum. In addition, the inverse $\mathrm{SNR}$ is equal to the relative noise of the detected signal. The key quantity is therefore the relative noise of the detected signal, which can originate from several physical sources.

1) Laser amplitude noise (dominant contribution)

The maser response to laser power is generally nonlinear. However, the sensitivity estimate is derived for small fluctuations around a fixed operating point, where a first-order linear approximation is valid. At our maser operating point, the calibrated laser produces a photocurrent of 47.8 mA in the photodetector (behind an ~OD2 neutral density filter) with a measured noise of 1 μA, corresponding to relative noise of approximately $2 \cdot 10^{-5}$. This small variation justifies the linear approximation used to convert optical power noise into equivalent signal noise. We note that strong nonlinear behaviour occurs close to threshold, where the maser can switch on and off. In contrast, our measurements are performed well beyond threshold.

2) Temperature fluctuations

Temperature variations shift the cavity resonance and can therefore change the detected signal amplitude. The solid copper cavity is buffered with helium gas to stabilize the temperature, and the remaining maser drift occurs only on hour timescales. Consequently, these slow drifts (< mHz) lie below the measurement bandwidth and do not contribute to the quoted sensitivity. While the sensitivity in a 1 Hz bandwidth is therefore only weakly affected, we emphasize in the manuscript that improved frequency stabilization will be required to ensure long-term stability and accuracy.

3) Magnetic-field and microwave-electronics noise

The purpose of the sensitivity estimate reported here is to evaluate the readout-limited sensitivity of the maser. Environmental magnetic noise from the bias magnet and the laboratory environment (including ambient fields such as the Earth's magnetic field) is considered part of the magnetic field the sensor is probed with. A separation of these contributions using calibrated AC magnetic fields has been performed by 9, which is, however, beyond the scope of the present work.

Based on the measured laser noise, we estimate a magnetic-field sensitivity of approximately

$$\frac{\partial B}{\sqrt{\Delta f}} \approx \frac{\Delta B}{\mathrm{SNR}} \cdot \frac{1}{\sqrt{\Delta f}} \approx \frac{1\mu\mathrm{T}}{1/(2 \cdot 10^{-5})} \cdot \frac{1}{\sqrt{1\mathrm{Hz}}} = 20\ \frac{\mathrm{pT}}{\sqrt{Hz}}$$

## Supplementary References